\documentclass[%
aps,prl,superscriptaddress,twocolumn,showpacs,preprintnumbers,amsmath,amssymb
]{revtex4-2}
\usepackage[normalem]{ulem}
\usepackage{graphicx}
\usepackage{color}
\usepackage{bm}
\usepackage{hyperref}
\usepackage{mathcomp}

\usepackage{amsmath}
\usepackage{amssymb}
\usepackage{empheq}
\usepackage{float}
\usepackage{multirow}
\usepackage{cases}
\usepackage{mathrsfs}
\usepackage{braket}
\usepackage[version=4]{mhchem}  

\newcommand{\equalcontribution}{\textsuperscript{\textbullet}} 

\begin{document}


\title{Raman spectroscopy of van der Waals topological magnet GdGaI}

\author{Nan~Jiang\equalcontribution}
\email{nan.jiang@phys.sci.osaka-u.ac.jp}
\affiliation{Department of Physics, Graduate School of Science, The University of Osaka, Toyonaka, Osaka 560-0043, Japan}
\affiliation{Center for Spintronics Research Network, The University of Osaka, Toyonaka 560-8531, Japan}
\affiliation{Institute for Open and Transdisciplinary Research Initiatives, The University of Osaka, Suita 565-0871,Japan}
\author{Yijin~Zhang\equalcontribution}
\email{zhang.yijin@phys.s.u-tokyo.ac.jp}
\affiliation{Institute of Industrial Science, The University of Tokyo, Meguro, Tokyo 153-8505, Japan}
\affiliation{Department of Physics, The University of Tokyo, Bunkyo, Tokyo 113-0033, Japan}
\author{Yujie~Xia\equalcontribution}
\affiliation{Institute of Theoretical Physics, Chinese Academy of Sciences, Beijing, China}
\author{Tomo~Higashihara}
\affiliation{Department of Physics, Graduate School of Science, The University of Osaka, Toyonaka, Osaka 560-0043, Japan}
\author{Ryutaro~Okuma}
\affiliation{Institute for Solid State Physics, The University of Tokyo, Kashiwa, Chiba 277-8581, Japan}
\author{Jun-ichi~Yamaura}
\affiliation{Institute for Solid State Physics, The University of Tokyo, Kashiwa, Chiba 277-8581, Japan}
\author{Yoshinori~Okada}
\affiliation{Okinawa Institute of Science and Technology Graduate University, Okinawa 904-0495, Japan}
\author{Kenji~Watanabe}
\affiliation{Research Center for Electronic and Optical Materials, National Institute for Materials Science, 1-1 Namiki, Tsukuba 305-0044, Japan}
\author{Takashi~Taniguchi }
\affiliation{Research Center for Materials Nanoarchitectonics, National Institute for Materials Science, 1-1 Namiki, Tsukuba 305-0044, Japan}
\author{Tiantian~Zhang}
\email{ttzhang@itp.ac.cn}
\affiliation{Institute of Theoretical Physics, Chinese Academy of Sciences, Beijing, China}
\author{Tomoki~Machida}
\affiliation{Institute of Industrial Science, The University of Tokyo, Meguro, Tokyo 153-8505, Japan}
\author{Yasuhiro~Niimi}
\affiliation{Department of Physics, Graduate School of Science, The University of Osaka, Toyonaka, Osaka 560-0043, Japan}
\affiliation{Center for Spintronics Research Network, The University of Osaka, Toyonaka 560-8531, Japan}
\affiliation{Institute for Open and Transdisciplinary Research Initiatives, The University of Osaka, Suita 565-0871,Japan}

\date{\today}

\begin{abstract}
We report polarization-resolved Raman spectroscopy of a van der Waals compound \ce{GdGaI} that is a candidate for excitonic insulators. 
By combining the symmetry analysis with density functional theory calculations, we identify six Raman-active phonons. 
The spectra exhibit only the expected anharmonic hardening down to $4~\mathrm{K}$: no additional peaks, no soft modes, and no signatures of zone folding are observed. 
This result indicates that any lattice distortion is below our experimental sensitivity, supporting an electronically driven origin for the band reconstruction reported by ARPES rather than an electron--phonon-driven mechanism.
Moreover, we observe a pronounced circular dichroism of the $A_{1g}$ modes under an out-of-plane magnetic field. 
Based on symmetry considerations, we attribute this dichroic response to chiral $A_{1g}$ phonons with opposite angular momenta generated by spin--phonon coupling in the time-reversal-broken state.
The temperature evolution of the degree of circular polarization further suggests that circularly polarized Raman spectroscopy detects the emergence of short-range antiferromagnetic correlations.
Our results highlight \ce{GdGaI} as a promising platform in which excitonic order, magnetism, and circularly polarized phonons can be intertwined, and demonstrate that circular-polarization Raman provides a sensitive probe of spin–phonon coupling in excitonic systems.
\end{abstract}

\maketitle

\section{INTRODUCTION}
Excitonic insulators (EIs) are ordered states of bound electron–hole pairs that condense below a critical temperature, provided that the interband Coulomb (excitonic) interaction overcomes the bare band gap of a narrow-gap semiconductor or a band-overlapped semimetal~\cite{Mott01021961,Keldysh1965,CLOIZEAUX1965259,PhysRev.158.462,PhysRev.162.752,PhysRevLett.19.439,Halperin1968,Kaneko2024}.  
The transition to an EI phase is accompanied by a reconstruction of the single-particle band dispersion and the opening of a gap at the Fermi level.  
During the past decade, numerous compounds, such as \ce{TiSe2}~\cite{Cercellier2007,TiSe2_2,TiSe2_3}, \ce{Ta2NiSe5}~\cite{Ta2NiSe5_1,Ta2NiSe5_2,Ta2NiSe5_3,Ta2NiSe5_4,Ta2NiSe5_5}, \ce{WTe2}~\cite{WTe2_1,WTe2_2}, \ce{ZrTe2}~\cite{ZrTe2}, and \ce{HfTe2}~\cite{HfTe2}, have been proposed as EI candidates on the basis of angle-resolved photoemission spectroscopy (ARPES), momentum-resolved electron energy-loss, and ultrafast pump–probe spectroscopy.  
Similar band reconstructions, however, also occur in conventional charge-density-wave (CDW) systems, where a Fermi-surface instability driven by electron–phonon coupling produces a periodic lattice distortion that folds the Brillouin zone and opens an energy gap~\cite{Gruner1994,TiSe2_3,Ta2NiSe5_8}.  
To discriminate between an excitonic mechanism and a lattice-driven CDW, one needs an experimental probe that is exquisitely sensitive to structural distortions.  

Polarized Raman spectroscopy satisfies this requirement.  
In CDW materials, the transition is accompanied by the softening of specific phonon branches and also the appearance of amplitude and zone-folded modes~\cite{Sugai1985,Lin2020}.  
By contrast, a purely electronic EI transition, that does not rely on strong electron–phonon coupling, is expected to leave the phonon spectrum largely unchanged.

In addition to probing lattice distortions, Raman scattering has become a powerful tool for studying magnetism in two-dimensional (2D) van der Waals (vdW) crystals. 
It reveals spin–phonon coupling~\cite{Tian_2016,PhysRevB.99.214304,Kim_2019,PhysRevB.107.075421,Wang_2020}, structural phase transitions, and Brillouin-zone folding~\cite{Kim2019,Lee2016,Wang_2016,Kim_rev}.
Moreover, Raman spectroscopy under a magnetic field enables direct optical readout of magnetic phases in magnets such as \ce{CrI3}~\cite{Zhang2020,PhysRevX.10.011075,Huang2020,McCreary2020} and \ce{VI3}~\cite{Lyu2020}. 
It can also detect spin-wave excitations~\cite{Cenker2021,PhysRevX.10.011075}, providing complementary insight into spin–lattice coupling.

Beyond probing magnetic phases, circular polarization also grants access to phonon chirality. 
Recent theory and experiment have established that lattice vibrations can carry angular momentum and become chiral when inversion or time-reversal symmetry is broken~\cite{PhysRevLett.115.115502,Zhu_Science2018,zhang2022chiral,Ueda2023,Ishito2023,zhang2023weyl,zhang2025chirality,zhang2025advances}. In magnetic crystals, spin–phonon coupling induces phonons with an angular momentum that is odd in the magnetization~\cite{ZhangNiu_PRL2014}. When time-reversal symmetry is broken by magnetic order or an external field, counter-rotating phonon partners can split and produce circular-dichroic responses in Raman/infrared spectra~\cite{Jin_PNAS2020,Hernandez_SciAdv2023,PhysRevLett.134.196905,PhysRevLett.134.196906,zhang2025chirality}.

The layered vdW compound \ce{GdGaI}~\cite{ggi0,ggikaneko} (space group $P\bar{3}m1$; Fig.~1(a)) has recently attracted attention as a promising EI platform~\cite{okumaggi}. ARPES studies on bulk crystals reveal folded conduction bands and a gap opening at low temperature, reminiscent of excitonic order.  First-principles and model analyses indicate that itinerant \ce{Gd}\,$5d$ electrons are exchange-coupled to localized \ce{Gd}\,$4f$ moments, stabilizing a triple-$\mathbf{q}$ non-collinear antiferromagnetic order whose Hund-coupling may favor a spin-triplet EI ground state~\cite{okumaggi}.

\begin{figure}
\includegraphics[width=85mm]{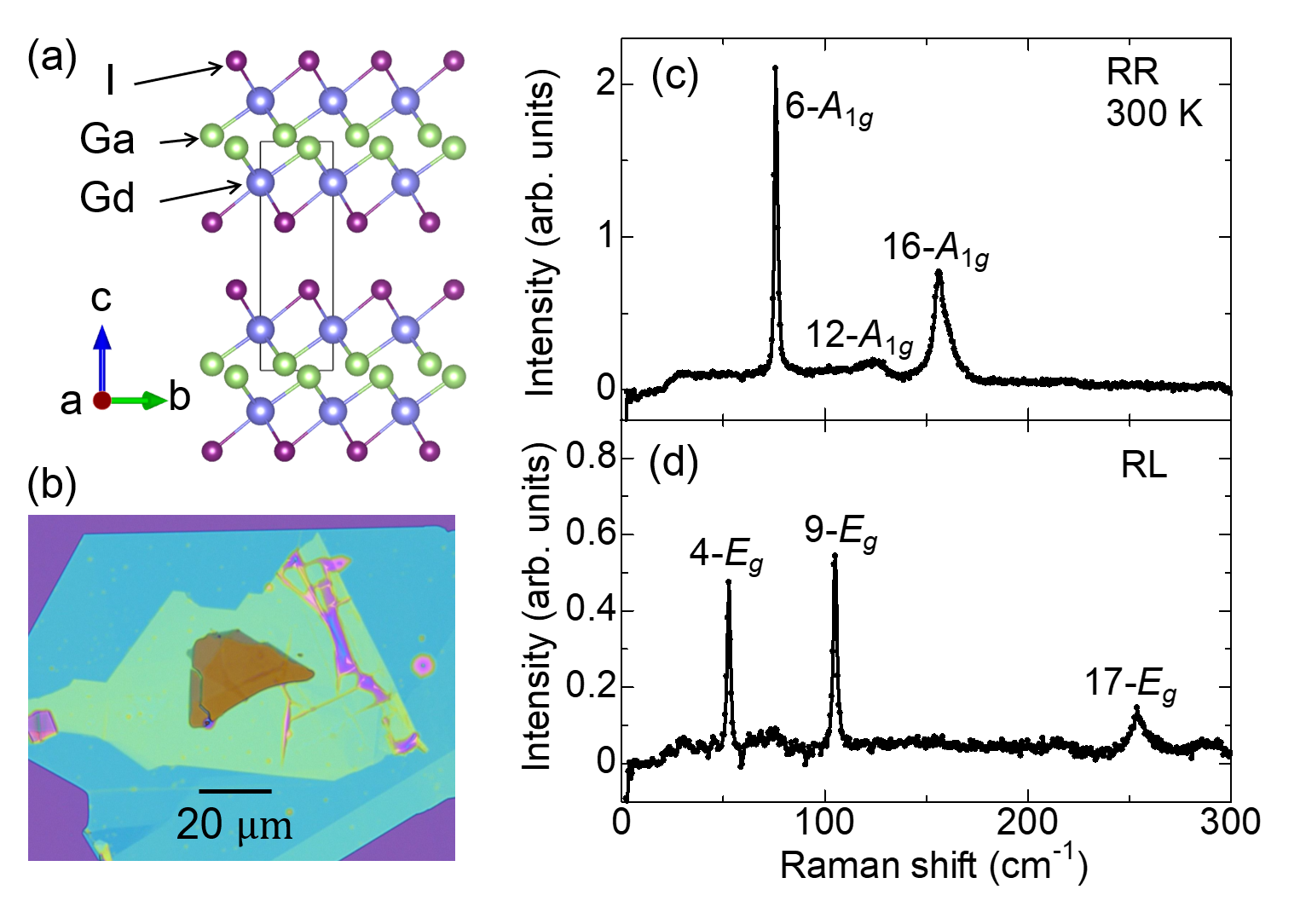}
\caption{\label{fig:epsart} (a) Crystal structure of GdGaI. The black solid line represents the unit cell of GdGaI. (b) An optical microscope image of GdGaI flake~No.~1 sandwiched by hBN flakes. (c,d) Polarized Raman spectra of the GdGaI flake No.~1 at 300~K measured with (c) RR and (d) RL scattering geometries.
}
\end{figure}

Here, we report a comprehensive polarized Raman investigation~\cite{Zhang2022} of exfoliated \ce{GdGaI} flakes down to \(T = 4~\mathrm{K}\) and under out-of-plane magnetic fields up to \(\mu_{0}H = \pm 9~\mathrm{T}\).  Density-functional-theory phonon calculations combined with the symmetry analysis allow us to assign all Raman-active modes at room temperature and to monitor their evolution with temperature and magnetic field. No additional phonon branches emerge and no mode exhibits anomalous softening on cooling, indicating that any CDW-type lattice instability is below our experimental sensitivity. These results support an interpretation that the band reconstruction in \ce{GdGaI} is driven predominantly by electron–electron interactions rather than electron–phonon coupling. Instead, \(A_{1g}\) modes display a clear magnetic-field dependence in the intensity ratio obtained with the incident/scattered right–right and left–left circular polarization configurations, consistent with circularly polarized phonons generated by spin–phonon coupling~\cite{sun2021review,wang2024chiral,zhang2025chirality,zhang2025general,zhang2025advances,juraschek2025chiral}. These results highlight \ce{GdGaI} as a rare system in which excitonic order, magnetism, and circularly polarized phonons can be intertwined.

\section{RAMAN SELECTION RULES AND PHONON ASSIGNMENT}
Figures~1(c) and 1(d) show polarized Raman spectra of \ce{GdGaI} acquired at $T = 300~\mathrm{K}$ with incident/scattered circular-polarization configurations of (c) right–right (RR) and (d) right–left (RL), respectively.  
A weak background signal, originating mainly from ambient \ce{N2} and minute organic contaminants at low Raman shifts, was measured on a substrate area lacking \ce{GdGaI} and subtracted from each spectrum (see Supplemental Materials Fig.~S2).  
After this correction, three phonon modes are clearly resolved in the polarization-conserving RR spectrum (Fig.~1(c)) and three distinct modes in the polarization-exchanging RL spectrum (Fig.~1(d)). Spatial mapping of Raman intensity confirm the uniformity of the flake as shown in Fig.~S3 in Supplemental Materials.

\begin{figure}
\includegraphics[width=85mm]{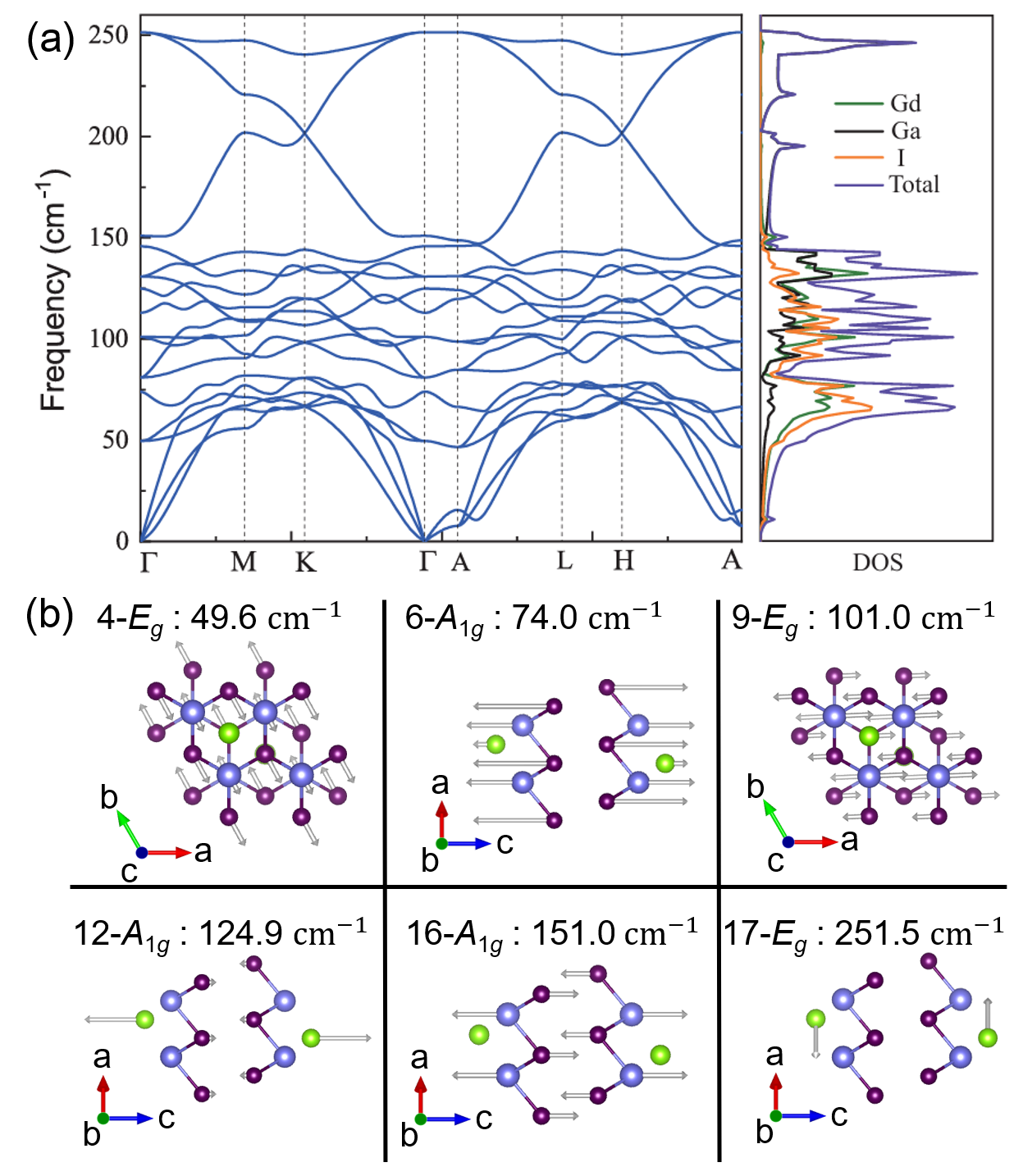}
\caption{\label{fig:epsart} (a) (a) Calculated phonon dispersion relations of \ce{GdGaI}, together with the total phonon density of states (DOS) and the atom-projected phonon DOS}. (b) Schematic illustration of Raman-active phonon modes in GdGaI at 300~K.
\end{figure}

The Raman scattering cross section is proportional to $\left|\mathbf{e}_{\mathrm{s}}^{\dagger} R \mathbf{e}_{\mathrm{i}}\right|^{2}$, where \(R\) is the Raman tensor and \(\mathbf{e}_{\mathrm{s}}\) and \(\mathbf{e}_{\mathrm{i}}\) are the polarization vectors of the scattered and incident light, respectively. 
In the paramagnetic state, \ce{GdGaI} belongs to the point group \(D_{3d}\) (\(\overline{3}m\)). 
The Raman tensors for the \(A_{1g}\) and \(E_{g}\) representations can be written as
\begin{align}
A_{1g} &=
\begin{pmatrix}
a & 0 & 0 \\
0 & a & 0 \\
0 & 0 & b
\end{pmatrix},
\label{eq:A1g}
\end{align}
and
\begin{subequations}\label{eq:Eg}
\begin{align}
E_{g}^{(1)} &=
\begin{pmatrix}
c & 0 & 0 \\
0 & -c & d \\
0 & d & 0
\end{pmatrix}, \label{eq:Eg1}\\[4pt]
E_{g}^{(2)} &=
\begin{pmatrix}
0 & -c & -d \\
-c & 0 & 0 \\
-d & 0 & 0
\end{pmatrix}, \label{eq:Eg2}
\end{align}
\end{subequations}
with real constants \(a\), \(b\), \(c\), and \(d\)~\cite{Loudon01101964}.

For circularly polarized light $\sigma^{\pm}=\frac{1}{\sqrt{2}}(1,\pm i,0)^{\mathsf T}$, the selection rules are as follows. 
The \(A_{1g}\) mode is active in the helicity-conserving configurations (RR and left-left (LL)), whereas the \(E_{g}\) modes are active in the helicity-exchanging configurations (RL and LR).

To assign the Raman peaks observed at room temperature, we have calculated the phonon spectra of \ce{GdGaI}. Details of the calculation are provided in the Supplemental Materials. Figure 2(a) displays the phonon dispersion along high-symmetry directions of the hexagonal Brillouin zone, together with the total phonon density of states (DOS) and the atom-projected phonon DOS.  
Starting from the lowest optical branch, we label the modes sequentially (numbers in Fig.~2) and identify six Raman-active modes: three non-degenerate $A_{1g}$ and three doubly degenerate $E_{g}$. The symmetries and frequencies of all phonon modes are summarized in Table~S1 of the Supplemental Materials.

Figure~2(b) show the vibration patterns for Raman-active modes.
The $A_{1g}$ modes involve out-of-plane breathing motions of the \ce{Ga}–\ce{I} layers relative to the \ce{Gd} sheet, while the $E_{g}$ modes correspond to in-plane shear and stretching vibrations.  
The calculated frequencies shown in Fig.~2(b) agree well with the peak positions obtained from the room-temperature Raman spectra in Figs.~1(c) and 1(d).

\begin{figure}
\includegraphics[width=75mm]{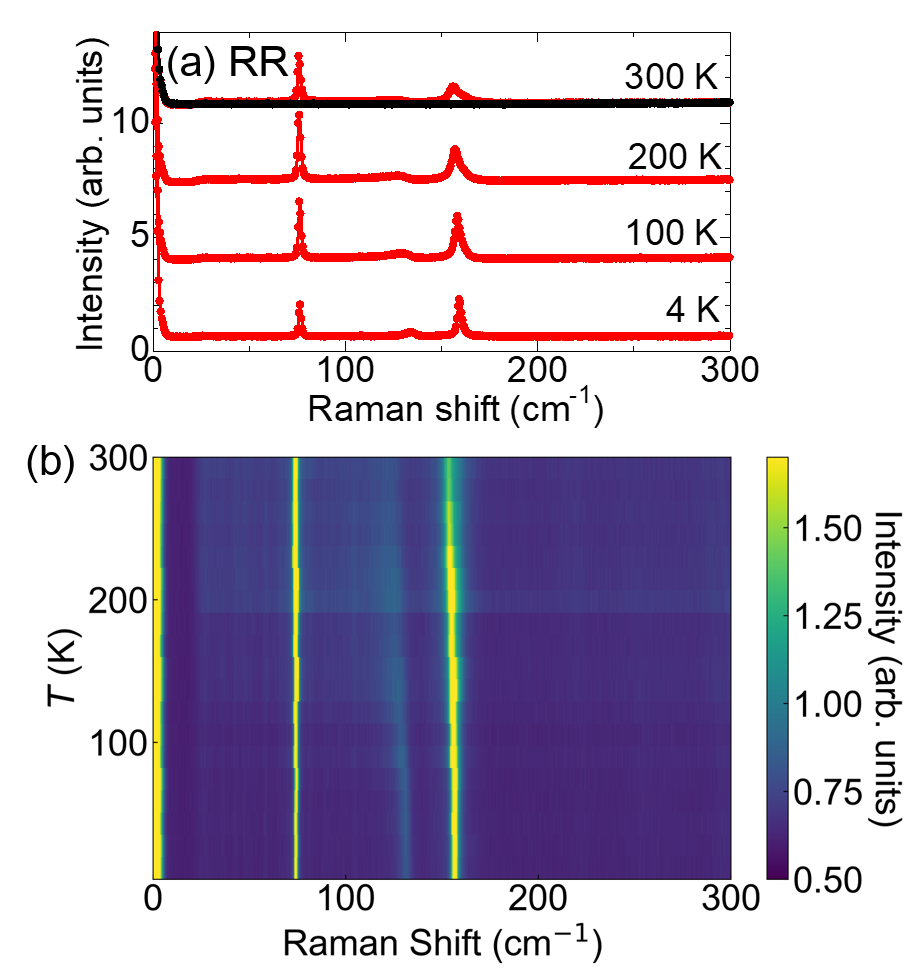}
\caption{\label{fig:epsart} (a) Polarized Raman spectra of GdGaI at several temperatures in the RR scattering geometry. The black trace represents the background signal measured on a substrate region devoid of GdGaI. (b) False-color map of the RR-polarized Raman response of GdGaI measured as a function of temperature.
}
\end{figure}

\section{TEMPERATURE DEPENDENCE OF RAMAN MODES}
To investigate the temperature dependence of phonon modes, we show a series of RR-polarized Raman spectra collected from $T = 300$ to $4~\mathrm{K}$ in Fig.~3(a).  
The black curve represents a background spectrum recorded at $300~\mathrm{K}$ on a region of the substrate devoid of \ce{GdGaI}.  
Previous ARPES measurements have shown that an excitonic gap starts to open below $\sim 200~\mathrm{K}$ and that a triple-$\mathbf{q}$ non-collinear antiferromagnetic order sets around $40~\mathrm{K}$~\cite{okumaggi}.  
If these transitions were driven by a periodic lattice distortion due to strong electron–phonon coupling, the Raman spectra would have shown marked changes. Typical signatures would have been phonon softening, anomalies in linewidth, and the appearance of zone-folded modes that reflect the new lattice periodicity.

In particular, formation of a $2\times 2\times 1$ superlattice would fold the electronic M point onto $\Gamma$, and phonons located at M would be folded to $\Gamma$ as well.
Our density-functional-theory calculations predict six $A_{g}$ and three $B_{g}$ modes at the M point below $300~\mathrm{cm^{-1}}$ (see Fig.~S4 in Supplemental Materials).  
Once folded, these modes would become Raman active and should be observable within the spectral window of our experiment.
Experimentally, however, no additional peaks emerge during cooling, and the line shapes of the three RR-active modes remain essentially unchanged down to $4~\mathrm{K}$.  
This absence of new Raman features is further illustrated in the false-colour intensity map of Fig.~3(b), which plots the RR spectra as a function of temperature.  
Equivalent measurements in the helicity-exchanged (RL) geometries (see Fig.~S5 in Supplemental Materials) show the same behavior.

\begin{figure}
\includegraphics[width=85mm]{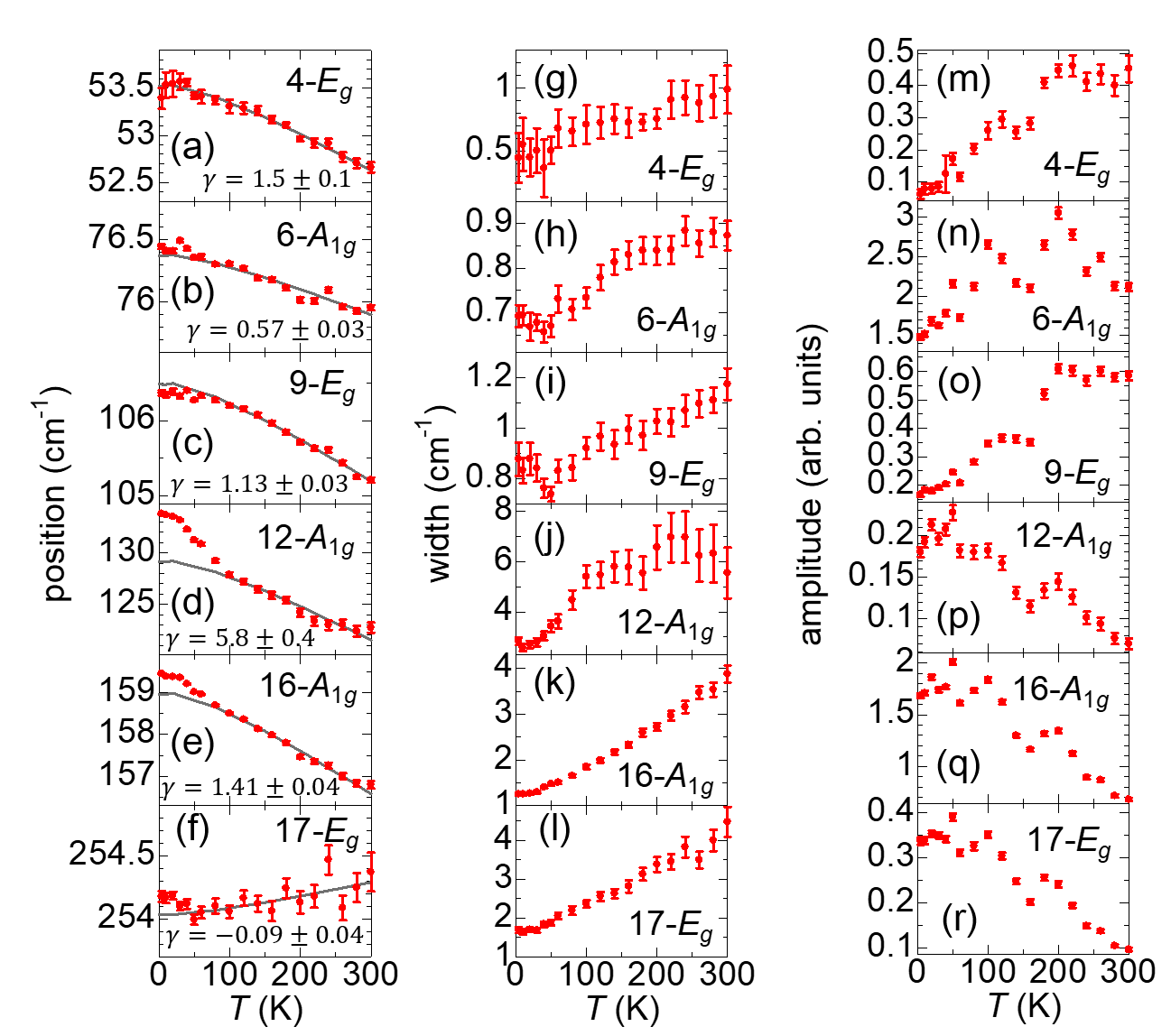}
\caption{\label{fig:epsart} Temperature dependence of the fitted (a–f) peak position, (g–l) full width at half maximum, and (m–r) amplitude for each Raman mode.  
All parameters were obtained from least-squares fits to Lorentzian profiles.  
Error bars denote $1\sigma$ uncertainties from the fit.
}
\end{figure}

To analyze the temperature evolution of the Raman modes, we performed Lorentzian fits to each peak at every temperature. Figures~4(a)--4(f), 4(g)--4(l), and 4(m)--4(r) summarize the extracted peak positions, linewidths, and amplitudes, respectively, for all observed modes. 
Before discussing these parameters, we clarify the uncertainties in the extracted quantities. Because the spectra were acquired over wide ranges of temperature and magnetic field, small variations in optical alignment and collection efficiency are unavoidable due to thermal contraction/expansion and magnetic forces. In practice, the optics must be realigned whenever the temperature or field is changed. As a result, the absolute intensity and, to a lesser extent, the linewidth can carry additional systematic uncertainty beyond the statistical errors returned by the Lorentzian fits (shown as error bars in Fig.~4). For this reason, we focus our discussion on the phonon frequencies (peak positions), which are most robust against such variations.

The temperature dependence of an optical-phonon frequency originates from anharmonic lattice vibrations and can be viewed as the sum of a contribution from lattice thermal expansion (or contraction) and a residual contribution that captures additional renormalizations beyond simple volume change.  Accordingly, we decompose the frequency shift relative to the harmonic-lattice value $\omega_{0}$ as
\begin{equation}
\Delta\omega(T) \equiv \omega(T)-\omega_{0}
= \Delta\omega_{\mathrm{TE}}(T) + \Delta\omega_{\mathrm{anh}}(T),
\label{eq:dw_decompose}
\end{equation}
where $\Delta\omega_{\mathrm{TE}}(T)$ accounts for the volume change of the lattice, while $\Delta\omega_{\mathrm{anh}}(T)$ summarizes all remaining effects, including intrinsic phonon--phonon anharmonicity and other interaction-driven contributions such as spin--phonon coupling~\cite{PhysRevB.29.2051,Irmer1996,Link1999}.  Within this framework, a smooth hardening upon cooling can be attributed primarily to lattice contraction and reduced anharmonic scattering, whereas sharp anomalies or extra modes would point to additional instabilities.

To quantify the thermal-expansion contribution in a simple manner, we adopt a Gr\"uneisen-type analysis in which $\Delta\omega_{\mathrm{TE}}(T)$ is parameterized solely by the fractional unit-cell volume change $\Delta V(T)/V$.  Using the measured lattice constants shown in Fig.~S6 in Supplemental Materials, we evaluate the unit-cell volume $V(T)$ and model the Raman peak frequency as
\begin{equation}
\omega(T)=\omega_{0}\left[1-\gamma\frac{V(T)-V_{\mathrm{300~K}}}{V_{\mathrm{300~K}}}\right],
\label{eq:gruneisen_fit}
\end{equation}
where $\omega_{0}$ is the mode frequency at $300~\mathrm{K}$, and $\gamma$ is an effective Gr\"uneisen parameter for each phonon mode, obtained from the temperature dependence of the phonon frequency under isobaric conditions~\cite{PhysRevB.12.3491,Ethan2019}. 
This parameter differs from the Gr\"uneisen parameter usually obtained from high-pressure Raman measurements under isothermal conditions, because it can include both the volume contribution due to thermal expansion and the intrinsic anharmonic contribution.
We determine $\omega_{0}$ and $\gamma$ by least-squares fits of Eq.~\eqref{eq:gruneisen_fit} over $100$--$300~\mathrm{K}$, and use the fitted parameters to generate the continuous curves plotted in Figs.~4(a)--4(f), with the fitted $\gamma$ annotated in each panel.

The high-temperature ($T\ge 100~\mathrm{K}$) phonon frequencies are well captured by this thermal-expansion trend. 
At lower temperatures, however, several modes---most notably the $12$--$A_{1g}$, $16$--$A_{1g}$, and $17$--$E_g$ modes---show systematic deviations from the Gr\"uneisen fit below $50$--$100~\mathrm{K}$. 
Given that \ce{GdGaI} undergoes magnetic ordering around $40~\mathrm{K}$~\cite{okumaggi}, we attribute these deviations to additional renormalizations beyond volume contraction, most plausibly magnetoelastic (spin--phonon) coupling, as widely observed in vdW magnetic materials~\cite{Tian_2016,PhysRevB.99.214304,Kim_2019,PhysRevB.107.075421,Wang_2020}. 
By contrast, the ARPES-reported band reconstruction onsets at much higher temperatures (around $200~\mathrm{K}$)~\cite{okumaggi}. We therefore associate the low-temperature deviations from the thermal-expansion trend with magnetic correlations, rather than with the onset of the band reconstruction. 
Taken together, these Raman results indicate that any CDW-type lattice distortion is below our experimental sensitivity and support an electronically driven origin of the band folding and gap opening reported by ARPES~\cite{okumaggi}.

\begin{figure}
\includegraphics[width=85mm]{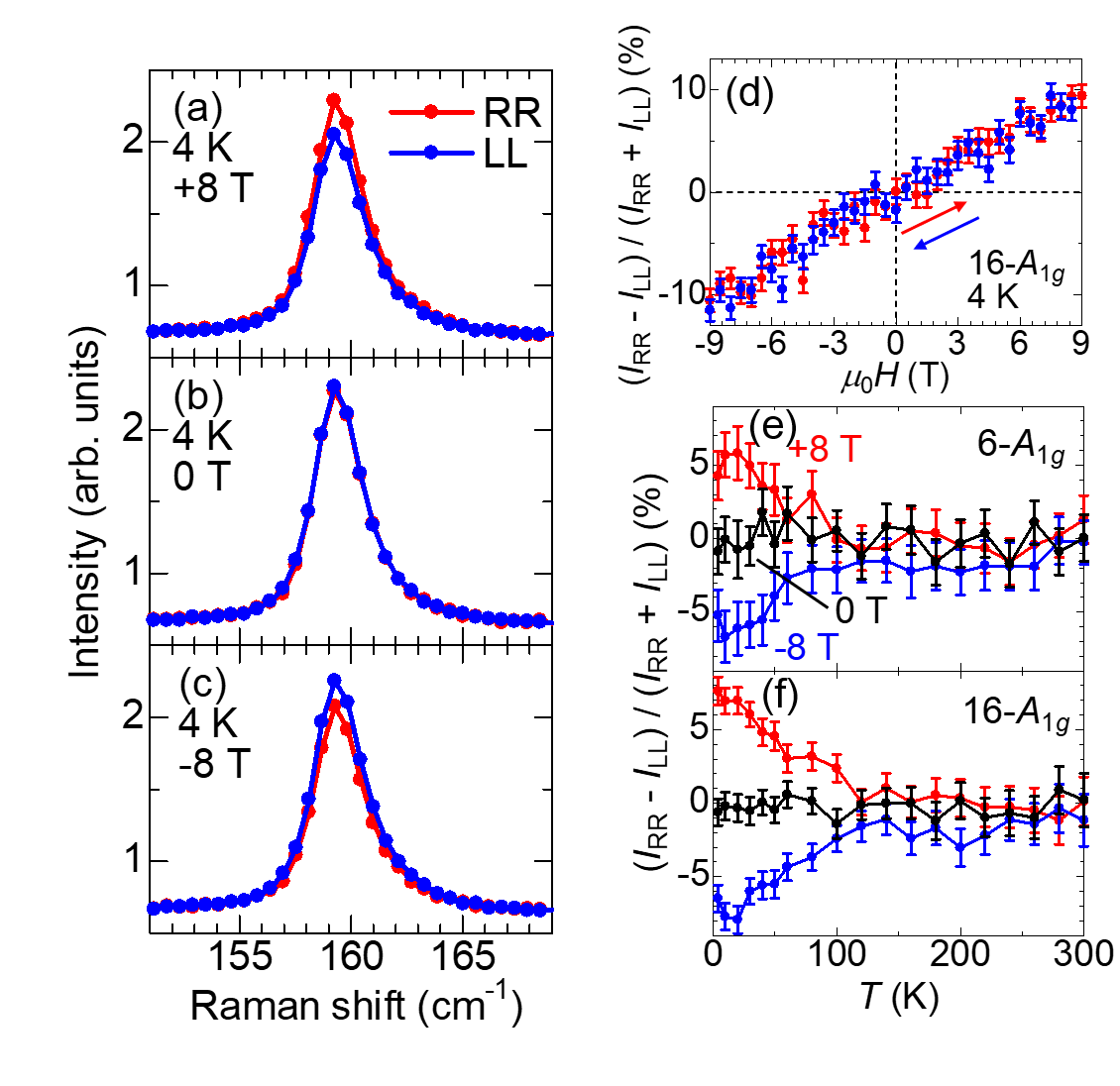}
\caption{\label{fig:epsart} (a-c) Polarized Raman spectra of the $16\text{-}A_{1g}$ phonon mode in GdGaI at 4~K, measured in RR (red) and LL (blue) scattering geometries under (a) 8~T, (b) 0~T, and (c) -8~T. 
(d) Magnetic‐field dependence of the degree of circular polarization at $T = 1.8~\mathrm{K}$ for the $16\text{-}A_{1g}$ phonon mode.
(e,f) Temperature dependence of the degree of circular polarization for the (e) $6\text{-}A_{1g}$ and (f) $16\text{-}A_{1g}$ modes under 8~T (red), 0~T (black), and -8~T (blue). Error bars denote $1\sigma$ uncertainties propagated from the Lorentzian fits.}
\end{figure}

\section{MAGNETIC-FIELD-INDUCED CIRCULAR DICHROISM}
Next, we show the magnetic field dependence of Raman spectrum. Figures~5(a)–5(c) compares the Raman spectra of the $16$–$A_{1g}$ mode measured at $T = 4~\mathrm{K}$ under magnetic fields of (a) $\mu_{0}H = +8~\mathrm{T}$, (b) $0~\mathrm{T}$, and (c) $-8~\mathrm{T}$.  
Red and blue symbols denote the RR and LL circular polarization configurations, respectively.  
The RR and LL intensities differ appreciably, and the sign of the difference reverses when the field direction is reversed. By contrast, the $E_{g}$ modes show no discernible difference between the RL and LR spectra under $\mu_{0}H=\pm 8~\mathrm{T}$ (see Fig.~S7 in Supplemental Materials).
This circular-dichroic response of the $A_{1g}$ mode stems from time-reversal symmetry breaking due to field-induced magnetization, which induces a finite angular momentum in the $A_{1g}$ phonon. The preserved $C_3$ symmetry forces this angular momentum to be parallel to the rotational axis~\cite{zhang2025chirality}, with a sign that follows the external field. This dependency results in the field-varying intensity presented in Figs. 5(a)-5(c). 

We now discuss the $A_{1g}$-mode intensity within the framework of the Raman tensor.
Similar magneto-optical Raman effects have been reported in van der Waals materials~\cite{Zhang2020,Huang2020,Lyu2020,adma.202101618,PhysRevB.108.184414,rao2025}.
In this context, an out‐of‐plane magnetic field reduces the symmetry from $\overline{3}m$ to $\overline{3}$ and introduces an antisymmetric imaginary component into the Raman tensor of the \(A_{1g}\) mode through the Faraday effect, yielding
\begin{align}
DA_{1g} &=
\begin{pmatrix}
  a   &  ie & 0 \\
 -ie  &  a  & 0 \\
  0   &  0  & b
\end{pmatrix},
\label{eq:DA1g}
\end{align}
where \(e\) is proportional to the magnetization~\cite{ANASTASSAKIS1972519,Lyu2020}.
Using Eq.~\eqref{eq:A1g}, the intensities in the RR and LL configurations are identical in the paramagnetic state,
\begin{equation}
I_{\mathrm{RR}} = I_{\mathrm{LL}} \propto
\left|(\sigma^{\pm})^{\dagger} R \sigma^{\pm}\right|^{2} = a^{2}.
\label{eq:IRR_paramag}
\end{equation} 
Under a field parallel to the $c$ axis, substitution of Eq.~\eqref{eq:DA1g} gives
\begin{subequations}
\begin{align}
I_{\mathrm{RR}} &\propto (a - e)^{2} \label{eq:IRR_field}\\
I_{\mathrm{LL}} &\propto (a + e)^{2} \label{eq:ILL_field}
\end{align}
\end{subequations}
which explain the field‐dependent intensity shown in Figs.~5(a)-5(c).

No difference is detected at $0~\mathrm{T}$ after sweeping the field from $+8~\mathrm{T}$ or from $-8~\mathrm{T}$ (see Figs.~S8(a) and S8(b) in Supplemental Materials), indicating that the remanent magnetization of \ce{GdGaI} is negligible compared with the one at $8~\mathrm{T}$. A small remanent magnetization attributed to canted ferromagnetism, reported for bulk crystals, is not resolved within our experimental sensitivity. 
We quantify the circular-polarization contrast by the degree of circular polarization of the Raman intensity, $(I_{\mathrm{RR}} - I_{\mathrm{LL}})/(I_{\mathrm{RR}} + I_{\mathrm{LL}})$, and we show its magnetic field dependence in Fig.~5(d).
It has the same tendency as the magnetization reported for bulk crystals~\cite{okumaggi}.
We have also confirmed the similar results for the $6\text{-}A_{1g}$ and $12\text{-}A_{1g}$ modes (see Fig.~S9 in Supplemental Materials). 

The same field-dependent behavior is observed for all $A_{1g}$ modes.
Figures~5(e) and 5(f) show the temperature dependence of the degree of circular polarization for the $6\text{-}A_{1g}$ and $16\text{-}A_{1g}$ modes, respectively.
Similar data for the $12\text{-}A_{1g}$ mode are provided in Fig.~S8(c) in Supplemental Materials.
The raw spectra of the $A_{1g}$ modes at each temperature and magnetic field are provided in Figs.~S10--S12 of the Supplemental Material.
The degree of circular polarization emerges below about $100~\mathrm{K}$ and increases rapidly below the bulk magnetic transition at approximately $40~\mathrm{K}$.
Notably, this onset temperature is consistent with the temperature range where several phonon frequencies begin to deviate from the thermal-expansion trend in our Gr\"uneisen analysis.
Bulk susceptibility measurements also show a comparable deviation from the Curie--Weiss law below $100~\mathrm{K}$~\cite{okumaggi}.
These observations indicate that the degree of circular polarization is linked to the development of short-range antiferromagnetic correlations in \ce{GdGaI}.

For comparison, we briefly comment on the $E_g$ modes.
The $E_g$ mode is doubly degenerate, with its constituent phonons possessing opposite angular momentum. Strong spin-phonon coupling can split this degeneracy, producing a circular-dichroic RL/LR splitting~\cite{Ishito2023,zhang2023weyl,PhysRevLett.134.196905,PhysRevLett.134.196906}. However, in \ce{GdGaI}, we observe no resolvable splitting within the experimental resolution between the RL and LR configurations.
By contrast, the $A_{1g}$ modes are non-degenerate and thus do not admit an $E_g$-like RL/LR splitting. The RR/LL intensity asymmetry observed for the $A_{1g}$ modes can arise from a field-induced antisymmetric (imaginary) Raman-tensor component, while the absence of a resolved RL/LR splitting for the $E_g$ modes indicates that any lifting of the $E_g$ degeneracy is below our experimental sensitivity.

\section{CONCLUSION}
In conclusion, we have performed polarized Raman spectroscopy on exfoliated \ce{GdGaI} flakes and identified six Raman–active phonons by combining the symmetry analysis with density–functional perturbation theory. The spectra exhibit only the expected anharmonic hardening and show no additional peaks or soft modes down to $4~\mathrm{K}$. These observations indicate that lattice distortion is negligible within our experimental sensitivity. The band folding and gap opening reported by ARPES are therefore attributed to electron–electron interactions rather than to a charge–density–wave mechanism driven by electron–phonon coupling.
Moreover, under out–of–plane magnetic fields up to $\mu_{0}H=\pm 9~\mathrm{T}$ we have observed a clear circular-dichroic response of the $A_{1g}$ Raman modes. This RR/LL intensity contrast is consistent with circularly polarized phonons that carry finite angular momentum generated by spin–phonon coupling in the time–reversal–broken state.
The temperature dependence of the degree of circular polarization indicates that the Raman intensity asymmetry is linked to the development of short-range antiferromagnetic correlations in \ce{GdGaI}. 
These results highlight \ce{GdGaI} as a platform in which excitonic order, magnetism, and circularly polarized phonons can be intertwined. 
We also demonstrate that polarization-resolved Raman spectroscopy provides a sensitive probe of spin–phonon coupling in excitonic-insulator systems.

The authors thank T. Kaneko and M. Ochi for fruitful discussions. The crystal structure of GdGaI was visualized using VESTA~\cite{vesta}. This work was supported by JSPS KAKENHI (Grant Nos. JP25K17347, JP22K18317, JP24K01293, JP21H05233 and JP23H02052), JST FOREST (Grant No. JPMJFR2134), JST PRESTO (Grant No. JPMJPR20L5, JPMJPR24H8), JST CREST (JPMJCR24A5), National Key R\&D Young Scientist Project (Grant Nos. 2023YFA1407400 and 2024YFA1400036), and the National Natural Science Foundation of China (Grant No. 12374165). K.W. and T.T. acknowledge support from World Premier International Research Center Initiative (WPI), MEXT, Japan.


\begin{thebibliography}{71}%
\makeatletter
\providecommand \@ifxundefined [1]{%
 \@ifx{#1\undefined}
}%
\providecommand \@ifnum [1]{%
 \ifnum #1\expandafter \@firstoftwo
 \else \expandafter \@secondoftwo
 \fi
}%
\providecommand \@ifx [1]{%
 \ifx #1\expandafter \@firstoftwo
 \else \expandafter \@secondoftwo
 \fi
}%
\providecommand \natexlab [1]{#1}%
\providecommand \enquote  [1]{``#1''}%
\providecommand \bibnamefont  [1]{#1}%
\providecommand \bibfnamefont [1]{#1}%
\providecommand \citenamefont [1]{#1}%
\providecommand \href@noop [0]{\@secondoftwo}%
\providecommand \href [0]{\begingroup \@sanitize@url \@href}%
\providecommand \@href[1]{\@@startlink{#1}\@@href}%
\providecommand \@@href[1]{\endgroup#1\@@endlink}%
\providecommand \@sanitize@url [0]{\catcode `\\12\catcode `\$12\catcode `\&12\catcode `\#12\catcode `\^12\catcode `\_12\catcode `\%12\relax}%
\providecommand \@@startlink[1]{}%
\providecommand \@@endlink[0]{}%
\providecommand \url  [0]{\begingroup\@sanitize@url \@url }%
\providecommand \@url [1]{\endgroup\@href {#1}{\urlprefix }}%
\providecommand \urlprefix  [0]{URL }%
\providecommand \Eprint [0]{\href }%
\providecommand \doibase [0]{https://doi.org/}%
\providecommand \selectlanguage [0]{\@gobble}%
\providecommand \bibinfo  [0]{\@secondoftwo}%
\providecommand \bibfield  [0]{\@secondoftwo}%
\providecommand \translation [1]{[#1]}%
\providecommand \BibitemOpen [0]{}%
\providecommand \bibitemStop [0]{}%
\providecommand \bibitemNoStop [0]{.\EOS\space}%
\providecommand \EOS [0]{\spacefactor3000\relax}%
\providecommand \BibitemShut  [1]{\csname bibitem#1\endcsname}%
\let\auto@bib@innerbib\@empty
\bibitem [{\citenamefont {Mott}(1961)}]{Mott01021961}%
  \BibitemOpen
  \bibfield  {author} {\bibinfo {author} {\bibfnamefont {N.~F.}\ \bibnamefont {Mott}},\ }\bibfield  {title} {\bibinfo {title} {{The transition to the metallic state}},\ }\href {https://doi.org/10.1080/14786436108243318} {\bibfield  {journal} {\bibinfo  {journal} {The Philosophical Magazine: A Journal of Theoretical Experimental and Applied Physics}\ }\textbf {\bibinfo {volume} {6}},\ \bibinfo {pages} {287} (\bibinfo {year} {1961})}\BibitemShut {NoStop}%
\bibitem [{\citenamefont {Keldysh}\ and\ \citenamefont {Kopaev}(1965)}]{Keldysh1965}%
  \BibitemOpen
  \bibfield  {author} {\bibinfo {author} {\bibfnamefont {L.~V.}\ \bibnamefont {Keldysh}}\ and\ \bibinfo {author} {\bibfnamefont {Y.~V.}\ \bibnamefont {Kopaev}},\ }\bibfield  {title} {\bibinfo {title} {{Possible instability of semimetallic state toward Coulomb interaction.}},\ }\href@noop {} {\bibfield  {journal} {\bibinfo  {journal} {Sov. Phys. Solid State}\ }\textbf {\bibinfo {volume} {6}},\ \bibinfo {pages} {2219} (\bibinfo {year} {1965})}\BibitemShut {NoStop}%
\bibitem [{\citenamefont {Cloizeaux}(1965)}]{CLOIZEAUX1965259}%
  \BibitemOpen
  \bibfield  {author} {\bibinfo {author} {\bibfnamefont {J.~D.}\ \bibnamefont {Cloizeaux}},\ }\bibfield  {title} {\bibinfo {title} {{Exciton instability and crystallographic anomalies in semiconductors}},\ }\href {https://doi.org/https://doi.org/10.1016/0022-3697(65)90153-8} {\bibfield  {journal} {\bibinfo  {journal} {Journal of Physics and Chemistry of Solids}\ }\textbf {\bibinfo {volume} {26}},\ \bibinfo {pages} {259} (\bibinfo {year} {1965})}\BibitemShut {NoStop}%
\bibitem [{\citenamefont {J\'erome}\ \emph {et~al.}(1967)\citenamefont {J\'erome}, \citenamefont {Rice},\ and\ \citenamefont {Kohn}}]{PhysRev.158.462}%
  \BibitemOpen
  \bibfield  {author} {\bibinfo {author} {\bibfnamefont {D.}~\bibnamefont {J\'erome}}, \bibinfo {author} {\bibfnamefont {T.~M.}\ \bibnamefont {Rice}},\ and\ \bibinfo {author} {\bibfnamefont {W.}~\bibnamefont {Kohn}},\ }\bibfield  {title} {\bibinfo {title} {{Excitonic Insulator}},\ }\href {https://doi.org/10.1103/PhysRev.158.462} {\bibfield  {journal} {\bibinfo  {journal} {Phys. Rev.}\ }\textbf {\bibinfo {volume} {158}},\ \bibinfo {pages} {462} (\bibinfo {year} {1967})}\BibitemShut {NoStop}%
\bibitem [{\citenamefont {Zittartz}(1967)}]{PhysRev.162.752}%
  \BibitemOpen
  \bibfield  {author} {\bibinfo {author} {\bibfnamefont {J.}~\bibnamefont {Zittartz}},\ }\bibfield  {title} {\bibinfo {title} {{Anisotropy Effects in the Excitonic Insulator}},\ }\href {https://doi.org/10.1103/PhysRev.162.752} {\bibfield  {journal} {\bibinfo  {journal} {Phys. Rev.}\ }\textbf {\bibinfo {volume} {162}},\ \bibinfo {pages} {752} (\bibinfo {year} {1967})}\BibitemShut {NoStop}%
\bibitem [{\citenamefont {Kohn}(1967)}]{PhysRevLett.19.439}%
  \BibitemOpen
  \bibfield  {author} {\bibinfo {author} {\bibfnamefont {W.}~\bibnamefont {Kohn}},\ }\bibfield  {title} {\bibinfo {title} {{Excitonic Phases}},\ }\href {https://doi.org/10.1103/PhysRevLett.19.439} {\bibfield  {journal} {\bibinfo  {journal} {Phys. Rev. Lett.}\ }\textbf {\bibinfo {volume} {19}},\ \bibinfo {pages} {439} (\bibinfo {year} {1967})}\BibitemShut {NoStop}%
\bibitem [{\citenamefont {HALPERIN}\ and\ \citenamefont {RICE}(1968)}]{Halperin1968}%
  \BibitemOpen
  \bibfield  {author} {\bibinfo {author} {\bibfnamefont {B.~I.}\ \bibnamefont {HALPERIN}}\ and\ \bibinfo {author} {\bibfnamefont {T.~M.}\ \bibnamefont {RICE}},\ }\bibfield  {title} {\bibinfo {title} {{Possible Anomalies at a Semimetal-Semiconductor Transistion}},\ }\href {https://doi.org/10.1103/RevModPhys.40.755} {\bibfield  {journal} {\bibinfo  {journal} {Rev. Mod. Phys.}\ }\textbf {\bibinfo {volume} {40}},\ \bibinfo {pages} {755} (\bibinfo {year} {1968})}\BibitemShut {NoStop}%
\bibitem [{\citenamefont {Kaneko}\ and\ \citenamefont {Ohta}(2025)}]{Kaneko2024}%
  \BibitemOpen
  \bibfield  {author} {\bibinfo {author} {\bibfnamefont {T.}~\bibnamefont {Kaneko}}\ and\ \bibinfo {author} {\bibfnamefont {Y.}~\bibnamefont {Ohta}},\ }\bibfield  {title} {\bibinfo {title} {{A New Era of Excitonic Insulators}},\ }\href {https://doi.org/10.7566/JPSJ.94.012001} {\bibfield  {journal} {\bibinfo  {journal} {Journal of the Physical Society of Japan}\ }\textbf {\bibinfo {volume} {94}},\ \bibinfo {pages} {012001} (\bibinfo {year} {2025})}\BibitemShut {NoStop}%
\bibitem [{\citenamefont {Cercellier}\ \emph {et~al.}(2007)\citenamefont {Cercellier}, \citenamefont {Monney}, \citenamefont {Clerc}, \citenamefont {Battaglia}, \citenamefont {Despont}, \citenamefont {Garnier}, \citenamefont {Beck}, \citenamefont {Aebi}, \citenamefont {Patthey}, \citenamefont {Berger},\ and\ \citenamefont {Forr\'o}}]{Cercellier2007}%
  \BibitemOpen
  \bibfield  {author} {\bibinfo {author} {\bibfnamefont {H.}~\bibnamefont {Cercellier}}, \bibinfo {author} {\bibfnamefont {C.}~\bibnamefont {Monney}}, \bibinfo {author} {\bibfnamefont {F.}~\bibnamefont {Clerc}}, \bibinfo {author} {\bibfnamefont {C.}~\bibnamefont {Battaglia}}, \bibinfo {author} {\bibfnamefont {L.}~\bibnamefont {Despont}}, \bibinfo {author} {\bibfnamefont {M.~G.}\ \bibnamefont {Garnier}}, \bibinfo {author} {\bibfnamefont {H.}~\bibnamefont {Beck}}, \bibinfo {author} {\bibfnamefont {P.}~\bibnamefont {Aebi}}, \bibinfo {author} {\bibfnamefont {L.}~\bibnamefont {Patthey}}, \bibinfo {author} {\bibfnamefont {H.}~\bibnamefont {Berger}},\ and\ \bibinfo {author} {\bibfnamefont {L.}~\bibnamefont {Forr\'o}},\ }\bibfield  {title} {\bibinfo {title} {{Evidence for an Excitonic Insulator Phase in $1T\mathrm{\text{\ensuremath{-}}}{\mathrm{TiSe}}_{2}$}},\ }\href {https://doi.org/10.1103/PhysRevLett.99.146403} {\bibfield  {journal} {\bibinfo  {journal} {Phys. Rev. Lett.}\ }\textbf {\bibinfo {volume} {99}},\
  \bibinfo {pages} {146403} (\bibinfo {year} {2007})}\BibitemShut {NoStop}%
\bibitem [{\citenamefont {Kogar}\ \emph {et~al.}(2017)\citenamefont {Kogar}, \citenamefont {Rak}, \citenamefont {Vig}, \citenamefont {Husain}, \citenamefont {Flicker}, \citenamefont {Joe}, \citenamefont {Venema}, \citenamefont {MacDougall}, \citenamefont {Chiang}, \citenamefont {Fradkin}, \citenamefont {van Wezel},\ and\ \citenamefont {Abbamonte}}]{TiSe2_2}%
  \BibitemOpen
  \bibfield  {author} {\bibinfo {author} {\bibfnamefont {A.}~\bibnamefont {Kogar}}, \bibinfo {author} {\bibfnamefont {M.~S.}\ \bibnamefont {Rak}}, \bibinfo {author} {\bibfnamefont {S.}~\bibnamefont {Vig}}, \bibinfo {author} {\bibfnamefont {A.~A.}\ \bibnamefont {Husain}}, \bibinfo {author} {\bibfnamefont {F.}~\bibnamefont {Flicker}}, \bibinfo {author} {\bibfnamefont {Y.~I.}\ \bibnamefont {Joe}}, \bibinfo {author} {\bibfnamefont {L.}~\bibnamefont {Venema}}, \bibinfo {author} {\bibfnamefont {G.~J.}\ \bibnamefont {MacDougall}}, \bibinfo {author} {\bibfnamefont {T.~C.}\ \bibnamefont {Chiang}}, \bibinfo {author} {\bibfnamefont {E.}~\bibnamefont {Fradkin}}, \bibinfo {author} {\bibfnamefont {J.}~\bibnamefont {van Wezel}},\ and\ \bibinfo {author} {\bibfnamefont {P.}~\bibnamefont {Abbamonte}},\ }\bibfield  {title} {\bibinfo {title} {{Signatures of exciton condensation in a transition metal dichalcogenide}},\ }\href {https://doi.org/10.1126/science.aam6432} {\bibfield  {journal} {\bibinfo  {journal} {Science}\ }\textbf
  {\bibinfo {volume} {358}},\ \bibinfo {pages} {1314} (\bibinfo {year} {2017})}\BibitemShut {NoStop}%
\bibitem [{\citenamefont {Holt}\ \emph {et~al.}(2001)\citenamefont {Holt}, \citenamefont {Zschack}, \citenamefont {Hong}, \citenamefont {Chou},\ and\ \citenamefont {Chiang}}]{TiSe2_3}%
  \BibitemOpen
  \bibfield  {author} {\bibinfo {author} {\bibfnamefont {M.}~\bibnamefont {Holt}}, \bibinfo {author} {\bibfnamefont {P.}~\bibnamefont {Zschack}}, \bibinfo {author} {\bibfnamefont {H.}~\bibnamefont {Hong}}, \bibinfo {author} {\bibfnamefont {M.~Y.}\ \bibnamefont {Chou}},\ and\ \bibinfo {author} {\bibfnamefont {T.-C.}\ \bibnamefont {Chiang}},\ }\bibfield  {title} {\bibinfo {title} {{X-Ray Studies of Phonon Softening in ${\mathrm{TiSe}}_{2}$}},\ }\href {https://doi.org/10.1103/PhysRevLett.86.3799} {\bibfield  {journal} {\bibinfo  {journal} {Phys. Rev. Lett.}\ }\textbf {\bibinfo {volume} {86}},\ \bibinfo {pages} {3799} (\bibinfo {year} {2001})}\BibitemShut {NoStop}%
\bibitem [{\citenamefont {Wakisaka}\ \emph {et~al.}(2009)\citenamefont {Wakisaka}, \citenamefont {Sudayama}, \citenamefont {Takubo}, \citenamefont {Mizokawa}, \citenamefont {Arita}, \citenamefont {Namatame}, \citenamefont {Taniguchi}, \citenamefont {Katayama}, \citenamefont {Nohara},\ and\ \citenamefont {Takagi}}]{Ta2NiSe5_1}%
  \BibitemOpen
  \bibfield  {author} {\bibinfo {author} {\bibfnamefont {Y.}~\bibnamefont {Wakisaka}}, \bibinfo {author} {\bibfnamefont {T.}~\bibnamefont {Sudayama}}, \bibinfo {author} {\bibfnamefont {K.}~\bibnamefont {Takubo}}, \bibinfo {author} {\bibfnamefont {T.}~\bibnamefont {Mizokawa}}, \bibinfo {author} {\bibfnamefont {M.}~\bibnamefont {Arita}}, \bibinfo {author} {\bibfnamefont {H.}~\bibnamefont {Namatame}}, \bibinfo {author} {\bibfnamefont {M.}~\bibnamefont {Taniguchi}}, \bibinfo {author} {\bibfnamefont {N.}~\bibnamefont {Katayama}}, \bibinfo {author} {\bibfnamefont {M.}~\bibnamefont {Nohara}},\ and\ \bibinfo {author} {\bibfnamefont {H.}~\bibnamefont {Takagi}},\ }\bibfield  {title} {\bibinfo {title} {{Excitonic Insulator State in ${\mathrm{Ta}}_{2}{\mathrm{NiSe}}_{5}$ Probed by Photoemission Spectroscopy}},\ }\href {https://doi.org/10.1103/PhysRevLett.103.026402} {\bibfield  {journal} {\bibinfo  {journal} {Phys. Rev. Lett.}\ }\textbf {\bibinfo {volume} {103}},\ \bibinfo {pages} {026402} (\bibinfo {year}
  {2009})}\BibitemShut {NoStop}%
\bibitem [{\citenamefont {Seki}\ \emph {et~al.}(2014)\citenamefont {Seki}, \citenamefont {Wakisaka}, \citenamefont {Kaneko}, \citenamefont {Toriyama}, \citenamefont {Konishi}, \citenamefont {Sudayama}, \citenamefont {Saini}, \citenamefont {Arita}, \citenamefont {Namatame}, \citenamefont {Taniguchi}, \citenamefont {Katayama}, \citenamefont {Nohara}, \citenamefont {Takagi}, \citenamefont {Mizokawa},\ and\ \citenamefont {Ohta}}]{Ta2NiSe5_2}%
  \BibitemOpen
  \bibfield  {author} {\bibinfo {author} {\bibfnamefont {K.}~\bibnamefont {Seki}}, \bibinfo {author} {\bibfnamefont {Y.}~\bibnamefont {Wakisaka}}, \bibinfo {author} {\bibfnamefont {T.}~\bibnamefont {Kaneko}}, \bibinfo {author} {\bibfnamefont {T.}~\bibnamefont {Toriyama}}, \bibinfo {author} {\bibfnamefont {T.}~\bibnamefont {Konishi}}, \bibinfo {author} {\bibfnamefont {T.}~\bibnamefont {Sudayama}}, \bibinfo {author} {\bibfnamefont {N.~L.}\ \bibnamefont {Saini}}, \bibinfo {author} {\bibfnamefont {M.}~\bibnamefont {Arita}}, \bibinfo {author} {\bibfnamefont {H.}~\bibnamefont {Namatame}}, \bibinfo {author} {\bibfnamefont {M.}~\bibnamefont {Taniguchi}}, \bibinfo {author} {\bibfnamefont {N.}~\bibnamefont {Katayama}}, \bibinfo {author} {\bibfnamefont {M.}~\bibnamefont {Nohara}}, \bibinfo {author} {\bibfnamefont {H.}~\bibnamefont {Takagi}}, \bibinfo {author} {\bibfnamefont {T.}~\bibnamefont {Mizokawa}},\ and\ \bibinfo {author} {\bibfnamefont {Y.}~\bibnamefont {Ohta}},\ }\bibfield  {title} {\bibinfo {title} {{Excitonic
  Bose-Einstein condensation in ${\mathrm{Ta}}_{2}{\mathrm{NiSe}}_{5}$ above room temperature}},\ }\href {https://doi.org/10.1103/PhysRevB.90.155116} {\bibfield  {journal} {\bibinfo  {journal} {Phys. Rev. B}\ }\textbf {\bibinfo {volume} {90}},\ \bibinfo {pages} {155116} (\bibinfo {year} {2014})}\BibitemShut {NoStop}%
\bibitem [{\citenamefont {Lu}\ \emph {et~al.}(2017)\citenamefont {Lu}, \citenamefont {Kono}, \citenamefont {Larkin}, \citenamefont {Rost}, \citenamefont {Takayama}, \citenamefont {Boris}, \citenamefont {Keimer},\ and\ \citenamefont {Takagi}}]{Ta2NiSe5_3}%
  \BibitemOpen
  \bibfield  {author} {\bibinfo {author} {\bibfnamefont {Y.~F.}\ \bibnamefont {Lu}}, \bibinfo {author} {\bibfnamefont {H.}~\bibnamefont {Kono}}, \bibinfo {author} {\bibfnamefont {T.~I.}\ \bibnamefont {Larkin}}, \bibinfo {author} {\bibfnamefont {A.~W.}\ \bibnamefont {Rost}}, \bibinfo {author} {\bibfnamefont {T.}~\bibnamefont {Takayama}}, \bibinfo {author} {\bibfnamefont {A.~V.}\ \bibnamefont {Boris}}, \bibinfo {author} {\bibfnamefont {B.}~\bibnamefont {Keimer}},\ and\ \bibinfo {author} {\bibfnamefont {H.}~\bibnamefont {Takagi}},\ }\bibfield  {title} {\bibinfo {title} {{Zero-gap semiconductor to excitonic insulator transition in Ta$_2$NiSe$_5$}},\ }\href {https://doi.org/10.1038/ncomms14408} {\bibfield  {journal} {\bibinfo  {journal} {Nature Communications}\ }\textbf {\bibinfo {volume} {8}},\ \bibinfo {pages} {14408} (\bibinfo {year} {2017})}\BibitemShut {NoStop}%
\bibitem [{\citenamefont {Werdehausen}\ \emph {et~al.}(2018)\citenamefont {Werdehausen}, \citenamefont {Takayama}, \citenamefont {Höppner}, \citenamefont {Albrecht}, \citenamefont {Rost}, \citenamefont {Lu}, \citenamefont {Manske}, \citenamefont {Takagi},\ and\ \citenamefont {Kaiser}}]{Ta2NiSe5_4}%
  \BibitemOpen
  \bibfield  {author} {\bibinfo {author} {\bibfnamefont {D.}~\bibnamefont {Werdehausen}}, \bibinfo {author} {\bibfnamefont {T.}~\bibnamefont {Takayama}}, \bibinfo {author} {\bibfnamefont {M.}~\bibnamefont {Höppner}}, \bibinfo {author} {\bibfnamefont {G.}~\bibnamefont {Albrecht}}, \bibinfo {author} {\bibfnamefont {A.~W.}\ \bibnamefont {Rost}}, \bibinfo {author} {\bibfnamefont {Y.}~\bibnamefont {Lu}}, \bibinfo {author} {\bibfnamefont {D.}~\bibnamefont {Manske}}, \bibinfo {author} {\bibfnamefont {H.}~\bibnamefont {Takagi}},\ and\ \bibinfo {author} {\bibfnamefont {S.}~\bibnamefont {Kaiser}},\ }\bibfield  {title} {\bibinfo {title} {{Coherent order parameter oscillations in the ground state of the excitonic insulator Ta$_2$NiSe$_5$}},\ }\href {https://doi.org/10.1126/sciadv.aap8652} {\bibfield  {journal} {\bibinfo  {journal} {Science Advances}\ }\textbf {\bibinfo {volume} {4}},\ \bibinfo {pages} {eaap8652} (\bibinfo {year} {2018})}\BibitemShut {NoStop}%
\bibitem [{\citenamefont {Mazza}\ \emph {et~al.}(2020)\citenamefont {Mazza}, \citenamefont {R\"osner}, \citenamefont {Windg\"atter}, \citenamefont {Latini}, \citenamefont {H\"ubener}, \citenamefont {Millis}, \citenamefont {Rubio},\ and\ \citenamefont {Georges}}]{Ta2NiSe5_5}%
  \BibitemOpen
  \bibfield  {author} {\bibinfo {author} {\bibfnamefont {G.}~\bibnamefont {Mazza}}, \bibinfo {author} {\bibfnamefont {M.}~\bibnamefont {R\"osner}}, \bibinfo {author} {\bibfnamefont {L.}~\bibnamefont {Windg\"atter}}, \bibinfo {author} {\bibfnamefont {S.}~\bibnamefont {Latini}}, \bibinfo {author} {\bibfnamefont {H.}~\bibnamefont {H\"ubener}}, \bibinfo {author} {\bibfnamefont {A.~J.}\ \bibnamefont {Millis}}, \bibinfo {author} {\bibfnamefont {A.}~\bibnamefont {Rubio}},\ and\ \bibinfo {author} {\bibfnamefont {A.}~\bibnamefont {Georges}},\ }\bibfield  {title} {\bibinfo {title} {{Nature of Symmetry Breaking at the Excitonic Insulator Transition: ${\mathrm{Ta}}_{2}{\mathrm{NiSe}}_{5}$}},\ }\href {https://doi.org/10.1103/PhysRevLett.124.197601} {\bibfield  {journal} {\bibinfo  {journal} {Phys. Rev. Lett.}\ }\textbf {\bibinfo {volume} {124}},\ \bibinfo {pages} {197601} (\bibinfo {year} {2020})}\BibitemShut {NoStop}%
\bibitem [{\citenamefont {Jia}\ \emph {et~al.}(2022)\citenamefont {Jia}, \citenamefont {Wang}, \citenamefont {Chiu}, \citenamefont {Song}, \citenamefont {Yu}, \citenamefont {J{\"a}ck}, \citenamefont {Lei}, \citenamefont {Klemenz}, \citenamefont {Cevallos}, \citenamefont {Onyszczak}, \citenamefont {Fishchenko}, \citenamefont {Liu}, \citenamefont {Farahi}, \citenamefont {Xie}, \citenamefont {Xu}, \citenamefont {Watanabe}, \citenamefont {Taniguchi}, \citenamefont {Bernevig}, \citenamefont {Cava}, \citenamefont {Schoop}, \citenamefont {Yazdani},\ and\ \citenamefont {Wu}}]{WTe2_1}%
  \BibitemOpen
  \bibfield  {author} {\bibinfo {author} {\bibfnamefont {Y.}~\bibnamefont {Jia}}, \bibinfo {author} {\bibfnamefont {P.}~\bibnamefont {Wang}}, \bibinfo {author} {\bibfnamefont {C.-L.}\ \bibnamefont {Chiu}}, \bibinfo {author} {\bibfnamefont {Z.}~\bibnamefont {Song}}, \bibinfo {author} {\bibfnamefont {G.}~\bibnamefont {Yu}}, \bibinfo {author} {\bibfnamefont {B.}~\bibnamefont {J{\"a}ck}}, \bibinfo {author} {\bibfnamefont {S.}~\bibnamefont {Lei}}, \bibinfo {author} {\bibfnamefont {S.}~\bibnamefont {Klemenz}}, \bibinfo {author} {\bibfnamefont {F.~A.}\ \bibnamefont {Cevallos}}, \bibinfo {author} {\bibfnamefont {M.}~\bibnamefont {Onyszczak}}, \bibinfo {author} {\bibfnamefont {N.}~\bibnamefont {Fishchenko}}, \bibinfo {author} {\bibfnamefont {X.}~\bibnamefont {Liu}}, \bibinfo {author} {\bibfnamefont {G.}~\bibnamefont {Farahi}}, \bibinfo {author} {\bibfnamefont {F.}~\bibnamefont {Xie}}, \bibinfo {author} {\bibfnamefont {Y.}~\bibnamefont {Xu}}, \bibinfo {author} {\bibfnamefont {K.}~\bibnamefont {Watanabe}}, \bibinfo
  {author} {\bibfnamefont {T.}~\bibnamefont {Taniguchi}}, \bibinfo {author} {\bibfnamefont {B.~A.}\ \bibnamefont {Bernevig}}, \bibinfo {author} {\bibfnamefont {R.~J.}\ \bibnamefont {Cava}}, \bibinfo {author} {\bibfnamefont {L.~M.}\ \bibnamefont {Schoop}}, \bibinfo {author} {\bibfnamefont {A.}~\bibnamefont {Yazdani}},\ and\ \bibinfo {author} {\bibfnamefont {S.}~\bibnamefont {Wu}},\ }\bibfield  {title} {\bibinfo {title} {{Evidence for a monolayer excitonic insulator}},\ }\href {https://doi.org/10.1038/s41567-021-01422-w} {\bibfield  {journal} {\bibinfo  {journal} {Nature Physics}\ }\textbf {\bibinfo {volume} {18}},\ \bibinfo {pages} {87} (\bibinfo {year} {2022})}\BibitemShut {NoStop}%
\bibitem [{\citenamefont {Sun}\ \emph {et~al.}(2022)\citenamefont {Sun}, \citenamefont {Zhao}, \citenamefont {Palomaki}, \citenamefont {Fei}, \citenamefont {Runburg}, \citenamefont {Malinowski}, \citenamefont {Huang}, \citenamefont {Cenker}, \citenamefont {Cui}, \citenamefont {Chu}, \citenamefont {Xu}, \citenamefont {Ataei}, \citenamefont {Varsano}, \citenamefont {Palummo}, \citenamefont {Molinari}, \citenamefont {Rontani},\ and\ \citenamefont {Cobden}}]{WTe2_2}%
  \BibitemOpen
  \bibfield  {author} {\bibinfo {author} {\bibfnamefont {B.}~\bibnamefont {Sun}}, \bibinfo {author} {\bibfnamefont {W.}~\bibnamefont {Zhao}}, \bibinfo {author} {\bibfnamefont {T.}~\bibnamefont {Palomaki}}, \bibinfo {author} {\bibfnamefont {Z.}~\bibnamefont {Fei}}, \bibinfo {author} {\bibfnamefont {E.}~\bibnamefont {Runburg}}, \bibinfo {author} {\bibfnamefont {P.}~\bibnamefont {Malinowski}}, \bibinfo {author} {\bibfnamefont {X.}~\bibnamefont {Huang}}, \bibinfo {author} {\bibfnamefont {J.}~\bibnamefont {Cenker}}, \bibinfo {author} {\bibfnamefont {Y.-T.}\ \bibnamefont {Cui}}, \bibinfo {author} {\bibfnamefont {J.-H.}\ \bibnamefont {Chu}}, \bibinfo {author} {\bibfnamefont {X.}~\bibnamefont {Xu}}, \bibinfo {author} {\bibfnamefont {S.~S.}\ \bibnamefont {Ataei}}, \bibinfo {author} {\bibfnamefont {D.}~\bibnamefont {Varsano}}, \bibinfo {author} {\bibfnamefont {M.}~\bibnamefont {Palummo}}, \bibinfo {author} {\bibfnamefont {E.}~\bibnamefont {Molinari}}, \bibinfo {author} {\bibfnamefont {M.}~\bibnamefont {Rontani}},\ and\
  \bibinfo {author} {\bibfnamefont {D.~H.}\ \bibnamefont {Cobden}},\ }\bibfield  {title} {\bibinfo {title} {{Evidence for equilibrium exciton condensation in monolayer WTe$_2$}},\ }\href {https://doi.org/10.1038/s41567-021-01427-5} {\bibfield  {journal} {\bibinfo  {journal} {Nature Physics}\ }\textbf {\bibinfo {volume} {18}},\ \bibinfo {pages} {94} (\bibinfo {year} {2022})}\BibitemShut {NoStop}%
\bibitem [{\citenamefont {Gao}\ \emph {et~al.}(2023)\citenamefont {Gao}, \citenamefont {Chan}, \citenamefont {Wang}, \citenamefont {Zhang}, \citenamefont {Jinxu}, \citenamefont {Cui}, \citenamefont {Yang}, \citenamefont {Liu}, \citenamefont {Shen}, \citenamefont {Sun}, \citenamefont {Jiang}, \citenamefont {Chiang},\ and\ \citenamefont {Chen}}]{ZrTe2}%
  \BibitemOpen
  \bibfield  {author} {\bibinfo {author} {\bibfnamefont {Q.}~\bibnamefont {Gao}}, \bibinfo {author} {\bibfnamefont {Y.-h.}\ \bibnamefont {Chan}}, \bibinfo {author} {\bibfnamefont {Y.}~\bibnamefont {Wang}}, \bibinfo {author} {\bibfnamefont {H.}~\bibnamefont {Zhang}}, \bibinfo {author} {\bibfnamefont {P.}~\bibnamefont {Jinxu}}, \bibinfo {author} {\bibfnamefont {S.}~\bibnamefont {Cui}}, \bibinfo {author} {\bibfnamefont {Y.}~\bibnamefont {Yang}}, \bibinfo {author} {\bibfnamefont {Z.}~\bibnamefont {Liu}}, \bibinfo {author} {\bibfnamefont {D.}~\bibnamefont {Shen}}, \bibinfo {author} {\bibfnamefont {Z.}~\bibnamefont {Sun}}, \bibinfo {author} {\bibfnamefont {J.}~\bibnamefont {Jiang}}, \bibinfo {author} {\bibfnamefont {T.~C.}\ \bibnamefont {Chiang}},\ and\ \bibinfo {author} {\bibfnamefont {P.}~\bibnamefont {Chen}},\ }\bibfield  {title} {\bibinfo {title} {{Evidence of high-temperature exciton condensation in a two-dimensional semimetal}},\ }\href {https://doi.org/10.1038/s41467-023-36667-x} {\bibfield  {journal}
  {\bibinfo  {journal} {Nature Communications}\ }\textbf {\bibinfo {volume} {14}},\ \bibinfo {pages} {994} (\bibinfo {year} {2023})}\BibitemShut {NoStop}%
\bibitem [{\citenamefont {Gao}\ \emph {et~al.}(2024)\citenamefont {Gao}, \citenamefont {Chan}, \citenamefont {Jiao}, \citenamefont {Chen}, \citenamefont {Yin}, \citenamefont {Tangprapha}, \citenamefont {Yang}, \citenamefont {Li}, \citenamefont {Liu}, \citenamefont {Shen}, \citenamefont {Jiang},\ and\ \citenamefont {Chen}}]{HfTe2}%
  \BibitemOpen
  \bibfield  {author} {\bibinfo {author} {\bibfnamefont {Q.}~\bibnamefont {Gao}}, \bibinfo {author} {\bibfnamefont {Y.-h.}\ \bibnamefont {Chan}}, \bibinfo {author} {\bibfnamefont {P.}~\bibnamefont {Jiao}}, \bibinfo {author} {\bibfnamefont {H.}~\bibnamefont {Chen}}, \bibinfo {author} {\bibfnamefont {S.}~\bibnamefont {Yin}}, \bibinfo {author} {\bibfnamefont {K.}~\bibnamefont {Tangprapha}}, \bibinfo {author} {\bibfnamefont {Y.}~\bibnamefont {Yang}}, \bibinfo {author} {\bibfnamefont {X.}~\bibnamefont {Li}}, \bibinfo {author} {\bibfnamefont {Z.}~\bibnamefont {Liu}}, \bibinfo {author} {\bibfnamefont {D.}~\bibnamefont {Shen}}, \bibinfo {author} {\bibfnamefont {S.}~\bibnamefont {Jiang}},\ and\ \bibinfo {author} {\bibfnamefont {P.}~\bibnamefont {Chen}},\ }\bibfield  {title} {\bibinfo {title} {{Observation of possible excitonic charge density waves and metal--insulator transitions in atomically thin semimetals}},\ }\href {https://doi.org/10.1038/s41567-023-02349-0} {\bibfield  {journal} {\bibinfo  {journal} {Nature
  Physics}\ }\textbf {\bibinfo {volume} {20}},\ \bibinfo {pages} {597} (\bibinfo {year} {2024})}\BibitemShut {NoStop}%
\bibitem [{\citenamefont {Gr{\"u}ner}(1994)}]{Gruner1994}%
  \BibitemOpen
  \bibfield  {author} {\bibinfo {author} {\bibfnamefont {G.}~\bibnamefont {Gr{\"u}ner}},\ }\href@noop {} {\emph {\bibinfo {title} {{Density Waves in Solids}}}},\ Frontiers in Physics\ (\bibinfo  {publisher} {Addison--Wesley},\ \bibinfo {address} {Reading, MA},\ \bibinfo {year} {1994})\BibitemShut {NoStop}%
\bibitem [{\citenamefont {Lu}\ \emph {et~al.}(2021)\citenamefont {Lu}, \citenamefont {Rossi}, \citenamefont {Kim}, \citenamefont {Yavas}, \citenamefont {Said}, \citenamefont {Nag}, \citenamefont {Garcia-Fernandez}, \citenamefont {Agrestini}, \citenamefont {Zhou}, \citenamefont {Jia}, \citenamefont {Moritz}, \citenamefont {Devereaux}, \citenamefont {Shen},\ and\ \citenamefont {Lee}}]{Ta2NiSe5_8}%
  \BibitemOpen
  \bibfield  {author} {\bibinfo {author} {\bibfnamefont {H.}~\bibnamefont {Lu}}, \bibinfo {author} {\bibfnamefont {M.}~\bibnamefont {Rossi}}, \bibinfo {author} {\bibfnamefont {J.-h.}\ \bibnamefont {Kim}}, \bibinfo {author} {\bibfnamefont {H.}~\bibnamefont {Yavas}}, \bibinfo {author} {\bibfnamefont {A.}~\bibnamefont {Said}}, \bibinfo {author} {\bibfnamefont {A.}~\bibnamefont {Nag}}, \bibinfo {author} {\bibfnamefont {M.}~\bibnamefont {Garcia-Fernandez}}, \bibinfo {author} {\bibfnamefont {S.}~\bibnamefont {Agrestini}}, \bibinfo {author} {\bibfnamefont {K.-J.}\ \bibnamefont {Zhou}}, \bibinfo {author} {\bibfnamefont {C.}~\bibnamefont {Jia}}, \bibinfo {author} {\bibfnamefont {B.}~\bibnamefont {Moritz}}, \bibinfo {author} {\bibfnamefont {T.~P.}\ \bibnamefont {Devereaux}}, \bibinfo {author} {\bibfnamefont {Z.-X.}\ \bibnamefont {Shen}},\ and\ \bibinfo {author} {\bibfnamefont {W.-S.}\ \bibnamefont {Lee}},\ }\bibfield  {title} {\bibinfo {title} {{Evolution of the electronic structure in
  ${\mathrm{Ta}}_{2}{\mathrm{NiSe}}_{5}$ across the structural transition revealed by resonant inelastic x-ray scattering}},\ }\href {https://doi.org/10.1103/PhysRevB.103.235159} {\bibfield  {journal} {\bibinfo  {journal} {Phys. Rev. B}\ }\textbf {\bibinfo {volume} {103}},\ \bibinfo {pages} {235159} (\bibinfo {year} {2021})}\BibitemShut {NoStop}%
\bibitem [{\citenamefont {Sugai}(1985)}]{Sugai1985}%
  \BibitemOpen
  \bibfield  {author} {\bibinfo {author} {\bibfnamefont {S.}~\bibnamefont {Sugai}},\ }\bibfield  {title} {\bibinfo {title} {{Lattice Vibrations in the Charge-Density-Wave States of Layered Transition Metal Dichalcogenides}},\ }\href {https://doi.org/https://doi.org/10.1002/pssb.2221290103} {\bibfield  {journal} {\bibinfo  {journal} {physica status solidi (b)}\ }\textbf {\bibinfo {volume} {129}},\ \bibinfo {pages} {13} (\bibinfo {year} {1985})}\BibitemShut {NoStop}%
\bibitem [{\citenamefont {Lin}\ \emph {et~al.}(2020)\citenamefont {Lin}, \citenamefont {Li}, \citenamefont {Wen}, \citenamefont {Berger}, \citenamefont {Forr{\'o}}, \citenamefont {Zhou}, \citenamefont {Jia}, \citenamefont {Taniguchi}, \citenamefont {Watanabe}, \citenamefont {Xi},\ and\ \citenamefont {Bahramy}}]{Lin2020}%
  \BibitemOpen
  \bibfield  {author} {\bibinfo {author} {\bibfnamefont {D.}~\bibnamefont {Lin}}, \bibinfo {author} {\bibfnamefont {S.}~\bibnamefont {Li}}, \bibinfo {author} {\bibfnamefont {J.}~\bibnamefont {Wen}}, \bibinfo {author} {\bibfnamefont {H.}~\bibnamefont {Berger}}, \bibinfo {author} {\bibfnamefont {L.}~\bibnamefont {Forr{\'o}}}, \bibinfo {author} {\bibfnamefont {H.}~\bibnamefont {Zhou}}, \bibinfo {author} {\bibfnamefont {S.}~\bibnamefont {Jia}}, \bibinfo {author} {\bibfnamefont {T.}~\bibnamefont {Taniguchi}}, \bibinfo {author} {\bibfnamefont {K.}~\bibnamefont {Watanabe}}, \bibinfo {author} {\bibfnamefont {X.}~\bibnamefont {Xi}},\ and\ \bibinfo {author} {\bibfnamefont {M.~S.}\ \bibnamefont {Bahramy}},\ }\bibfield  {title} {\bibinfo {title} {{Patterns and driving forces of dimensionality-dependent charge density waves in 2$H$-type transition metal dichalcogenides}},\ }\href {https://doi.org/10.1038/s41467-020-15715-w} {\bibfield  {journal} {\bibinfo  {journal} {Nature Communications}\ }\textbf {\bibinfo {volume}
  {11}},\ \bibinfo {pages} {2406} (\bibinfo {year} {2020})}\BibitemShut {NoStop}%
\bibitem [{\citenamefont {Tian}\ \emph {et~al.}(2016)\citenamefont {Tian}, \citenamefont {Gray}, \citenamefont {Ji}, \citenamefont {Cava},\ and\ \citenamefont {Burch}}]{Tian_2016}%
  \BibitemOpen
  \bibfield  {author} {\bibinfo {author} {\bibfnamefont {Y.}~\bibnamefont {Tian}}, \bibinfo {author} {\bibfnamefont {M.~J.}\ \bibnamefont {Gray}}, \bibinfo {author} {\bibfnamefont {H.}~\bibnamefont {Ji}}, \bibinfo {author} {\bibfnamefont {R.~J.}\ \bibnamefont {Cava}},\ and\ \bibinfo {author} {\bibfnamefont {K.~S.}\ \bibnamefont {Burch}},\ }\bibfield  {title} {\bibinfo {title} {{Magneto-elastic coupling in a potential ferromagnetic 2D atomic crystal}},\ }\href {https://doi.org/10.1088/2053-1583/3/2/025035} {\bibfield  {journal} {\bibinfo  {journal} {2D Materials}\ }\textbf {\bibinfo {volume} {3}},\ \bibinfo {pages} {025035} (\bibinfo {year} {2016})}\BibitemShut {NoStop}%
\bibitem [{\citenamefont {Milosavljevi\ifmmode~\acute{c}\else \'{c}\fi{}}\ \emph {et~al.}(2019)\citenamefont {Milosavljevi\ifmmode~\acute{c}\else \'{c}\fi{}}, \citenamefont {\ifmmode \check{S}\else \v{S}\fi{}olaji\ifmmode~\acute{c}\else \'{c}\fi{}}, \citenamefont {Djurdji\ifmmode \acute{c}\else~\'{c}\fi{} Mijin}, \citenamefont {Pe\ifmmode \check{s}\else \v{s}\fi{}i\ifmmode~\acute{c}\else \'{c}\fi{}}, \citenamefont {Vi\ifmmode \check{s}\else \v{s}\fi{}i\ifmmode~\acute{c}\else \'{c}\fi{}}, \citenamefont {Liu}, \citenamefont {Petrovic}, \citenamefont {Lazarevi\ifmmode~\acute{c}\else \'{c}\fi{}},\ and\ \citenamefont {Popovi\ifmmode~\acute{c}\else \'{c}\fi{}}}]{PhysRevB.99.214304}%
  \BibitemOpen
  \bibfield  {author} {\bibinfo {author} {\bibfnamefont {A.}~\bibnamefont {Milosavljevi\ifmmode~\acute{c}\else \'{c}\fi{}}}, \bibinfo {author} {\bibfnamefont {A.}~\bibnamefont {\ifmmode \check{S}\else \v{S}\fi{}olaji\ifmmode~\acute{c}\else \'{c}\fi{}}}, \bibinfo {author} {\bibfnamefont {S.}~\bibnamefont {Djurdji\ifmmode \acute{c}\else~\'{c}\fi{} Mijin}}, \bibinfo {author} {\bibfnamefont {J.}~\bibnamefont {Pe\ifmmode \check{s}\else \v{s}\fi{}i\ifmmode~\acute{c}\else \'{c}\fi{}}}, \bibinfo {author} {\bibfnamefont {B.}~\bibnamefont {Vi\ifmmode \check{s}\else \v{s}\fi{}i\ifmmode~\acute{c}\else \'{c}\fi{}}}, \bibinfo {author} {\bibfnamefont {Y.}~\bibnamefont {Liu}}, \bibinfo {author} {\bibfnamefont {C.}~\bibnamefont {Petrovic}}, \bibinfo {author} {\bibfnamefont {N.}~\bibnamefont {Lazarevi\ifmmode~\acute{c}\else \'{c}\fi{}}},\ and\ \bibinfo {author} {\bibfnamefont {Z.~V.}\ \bibnamefont {Popovi\ifmmode~\acute{c}\else \'{c}\fi{}}},\ }\bibfield  {title} {\bibinfo {title} {{Lattice dynamics and phase transitions in
  ${\mathrm{Fe}}_{3\ensuremath{-}x}{\mathrm{GeTe}}_{2}$}},\ }\href {https://doi.org/10.1103/PhysRevB.99.214304} {\bibfield  {journal} {\bibinfo  {journal} {Phys. Rev. B}\ }\textbf {\bibinfo {volume} {99}},\ \bibinfo {pages} {214304} (\bibinfo {year} {2019})}\BibitemShut {NoStop}%
\bibitem [{\citenamefont {Kim}\ \emph {et~al.}(2019{\natexlab{a}})\citenamefont {Kim}, \citenamefont {Lim}, \citenamefont {Kim}, \citenamefont {Lee}, \citenamefont {Lee}, \citenamefont {Kim}, \citenamefont {Park}, \citenamefont {Son}, \citenamefont {Park}, \citenamefont {Park},\ and\ \citenamefont {Cheong}}]{Kim_2019}%
  \BibitemOpen
  \bibfield  {author} {\bibinfo {author} {\bibfnamefont {K.}~\bibnamefont {Kim}}, \bibinfo {author} {\bibfnamefont {S.~Y.}\ \bibnamefont {Lim}}, \bibinfo {author} {\bibfnamefont {J.}~\bibnamefont {Kim}}, \bibinfo {author} {\bibfnamefont {J.-U.}\ \bibnamefont {Lee}}, \bibinfo {author} {\bibfnamefont {S.}~\bibnamefont {Lee}}, \bibinfo {author} {\bibfnamefont {P.}~\bibnamefont {Kim}}, \bibinfo {author} {\bibfnamefont {K.}~\bibnamefont {Park}}, \bibinfo {author} {\bibfnamefont {S.}~\bibnamefont {Son}}, \bibinfo {author} {\bibfnamefont {C.-H.}\ \bibnamefont {Park}}, \bibinfo {author} {\bibfnamefont {J.-G.}\ \bibnamefont {Park}},\ and\ \bibinfo {author} {\bibfnamefont {H.}~\bibnamefont {Cheong}},\ }\bibfield  {title} {\bibinfo {title} {{Antiferromagnetic ordering in van der Waals 2D magnetic material MnPS$_3$ probed by Raman spectroscopy}},\ }\href {https://doi.org/10.1088/2053-1583/ab27d5} {\bibfield  {journal} {\bibinfo  {journal} {2D Materials}\ }\textbf {\bibinfo {volume} {6}},\ \bibinfo {pages} {041001}
  (\bibinfo {year} {2019}{\natexlab{a}})}\BibitemShut {NoStop}%
\bibitem [{\citenamefont {Pawbake}\ \emph {et~al.}(2023)\citenamefont {Pawbake}, \citenamefont {Pelini}, \citenamefont {Wilson}, \citenamefont {Mosina}, \citenamefont {Sofer}, \citenamefont {Heid},\ and\ \citenamefont {Faugeras}}]{PhysRevB.107.075421}%
  \BibitemOpen
  \bibfield  {author} {\bibinfo {author} {\bibfnamefont {A.}~\bibnamefont {Pawbake}}, \bibinfo {author} {\bibfnamefont {T.}~\bibnamefont {Pelini}}, \bibinfo {author} {\bibfnamefont {N.~P.}\ \bibnamefont {Wilson}}, \bibinfo {author} {\bibfnamefont {K.}~\bibnamefont {Mosina}}, \bibinfo {author} {\bibfnamefont {Z.}~\bibnamefont {Sofer}}, \bibinfo {author} {\bibfnamefont {R.}~\bibnamefont {Heid}},\ and\ \bibinfo {author} {\bibfnamefont {C.}~\bibnamefont {Faugeras}},\ }\bibfield  {title} {\bibinfo {title} {{Raman scattering signatures of strong spin-phonon coupling in the bulk magnetic van der Waals material CrSBr}},\ }\href {https://doi.org/10.1103/PhysRevB.107.075421} {\bibfield  {journal} {\bibinfo  {journal} {Phys. Rev. B}\ }\textbf {\bibinfo {volume} {107}},\ \bibinfo {pages} {075421} (\bibinfo {year} {2023})}\BibitemShut {NoStop}%
\bibitem [{\citenamefont {Wang}\ \emph {et~al.}(2020)\citenamefont {Wang}, \citenamefont {Tian}, \citenamefont {Li}, \citenamefont {Jin}, \citenamefont {Ji}, \citenamefont {Lei},\ and\ \citenamefont {Zhang}}]{Wang_2020}%
  \BibitemOpen
  \bibfield  {author} {\bibinfo {author} {\bibfnamefont {Y.-M.}\ \bibnamefont {Wang}}, \bibinfo {author} {\bibfnamefont {S.-J.}\ \bibnamefont {Tian}}, \bibinfo {author} {\bibfnamefont {C.-H.}\ \bibnamefont {Li}}, \bibinfo {author} {\bibfnamefont {F.}~\bibnamefont {Jin}}, \bibinfo {author} {\bibfnamefont {J.-T.}\ \bibnamefont {Ji}}, \bibinfo {author} {\bibfnamefont {H.-C.}\ \bibnamefont {Lei}},\ and\ \bibinfo {author} {\bibfnamefont {Q.-M.}\ \bibnamefont {Zhang}},\ }\bibfield  {title} {\bibinfo {title} {{Raman scattering study of two-dimensional magnetic van der Waals compound VI$_3$}},\ }\href {https://doi.org/10.1088/1674-1056/ab8215} {\bibfield  {journal} {\bibinfo  {journal} {Chinese Physics B}\ }\textbf {\bibinfo {volume} {29}},\ \bibinfo {pages} {056301} (\bibinfo {year} {2020})}\BibitemShut {NoStop}%
\bibitem [{\citenamefont {Kim}\ \emph {et~al.}(2019{\natexlab{b}})\citenamefont {Kim}, \citenamefont {Lim}, \citenamefont {Lee}, \citenamefont {Lee}, \citenamefont {Kim}, \citenamefont {Park}, \citenamefont {Jeon}, \citenamefont {Park}, \citenamefont {Park},\ and\ \citenamefont {Cheong}}]{Kim2019}%
  \BibitemOpen
  \bibfield  {author} {\bibinfo {author} {\bibfnamefont {K.}~\bibnamefont {Kim}}, \bibinfo {author} {\bibfnamefont {S.~Y.}\ \bibnamefont {Lim}}, \bibinfo {author} {\bibfnamefont {J.-U.}\ \bibnamefont {Lee}}, \bibinfo {author} {\bibfnamefont {S.}~\bibnamefont {Lee}}, \bibinfo {author} {\bibfnamefont {T.~Y.}\ \bibnamefont {Kim}}, \bibinfo {author} {\bibfnamefont {K.}~\bibnamefont {Park}}, \bibinfo {author} {\bibfnamefont {G.~S.}\ \bibnamefont {Jeon}}, \bibinfo {author} {\bibfnamefont {C.-H.}\ \bibnamefont {Park}}, \bibinfo {author} {\bibfnamefont {J.-G.}\ \bibnamefont {Park}},\ and\ \bibinfo {author} {\bibfnamefont {H.}~\bibnamefont {Cheong}},\ }\bibfield  {title} {\bibinfo {title} {{Suppression of magnetic ordering in XXZ-type antiferromagnetic monolayer NiPS$_3$}},\ }\href {https://doi.org/10.1038/s41467-018-08284-6} {\bibfield  {journal} {\bibinfo  {journal} {Nature Communications}\ }\textbf {\bibinfo {volume} {10}},\ \bibinfo {pages} {345} (\bibinfo {year} {2019}{\natexlab{b}})}\BibitemShut {NoStop}%
\bibitem [{\citenamefont {Lee}\ \emph {et~al.}(2016)\citenamefont {Lee}, \citenamefont {Lee}, \citenamefont {Ryoo}, \citenamefont {Kang}, \citenamefont {Kim}, \citenamefont {Kim}, \citenamefont {Park}, \citenamefont {Park},\ and\ \citenamefont {Cheong}}]{Lee2016}%
  \BibitemOpen
  \bibfield  {author} {\bibinfo {author} {\bibfnamefont {J.-U.}\ \bibnamefont {Lee}}, \bibinfo {author} {\bibfnamefont {S.}~\bibnamefont {Lee}}, \bibinfo {author} {\bibfnamefont {J.~H.}\ \bibnamefont {Ryoo}}, \bibinfo {author} {\bibfnamefont {S.}~\bibnamefont {Kang}}, \bibinfo {author} {\bibfnamefont {T.~Y.}\ \bibnamefont {Kim}}, \bibinfo {author} {\bibfnamefont {P.}~\bibnamefont {Kim}}, \bibinfo {author} {\bibfnamefont {C.-H.}\ \bibnamefont {Park}}, \bibinfo {author} {\bibfnamefont {J.-G.}\ \bibnamefont {Park}},\ and\ \bibinfo {author} {\bibfnamefont {H.}~\bibnamefont {Cheong}},\ }\bibfield  {title} {\bibinfo {title} {{Ising-Type Magnetic Ordering in Atomically Thin FePS$_3$}},\ }\href {https://doi.org/10.1021/acs.nanolett.6b03052} {\bibfield  {journal} {\bibinfo  {journal} {Nano Letters}\ }\textbf {\bibinfo {volume} {16}},\ \bibinfo {pages} {7433} (\bibinfo {year} {2016})}\BibitemShut {NoStop}%
\bibitem [{\citenamefont {Wang}\ \emph {et~al.}(2016)\citenamefont {Wang}, \citenamefont {Du}, \citenamefont {Fredrik~Liu}, \citenamefont {Hu}, \citenamefont {Zhang}, \citenamefont {Zhang}, \citenamefont {Owen}, \citenamefont {Lu}, \citenamefont {Gan}, \citenamefont {Sengupta}, \citenamefont {Kloc},\ and\ \citenamefont {Xiong}}]{Wang_2016}%
  \BibitemOpen
  \bibfield  {author} {\bibinfo {author} {\bibfnamefont {X.}~\bibnamefont {Wang}}, \bibinfo {author} {\bibfnamefont {K.}~\bibnamefont {Du}}, \bibinfo {author} {\bibfnamefont {Y.~Y.}\ \bibnamefont {Fredrik~Liu}}, \bibinfo {author} {\bibfnamefont {P.}~\bibnamefont {Hu}}, \bibinfo {author} {\bibfnamefont {J.}~\bibnamefont {Zhang}}, \bibinfo {author} {\bibfnamefont {Q.}~\bibnamefont {Zhang}}, \bibinfo {author} {\bibfnamefont {M.~H.~S.}\ \bibnamefont {Owen}}, \bibinfo {author} {\bibfnamefont {X.}~\bibnamefont {Lu}}, \bibinfo {author} {\bibfnamefont {C.~K.}\ \bibnamefont {Gan}}, \bibinfo {author} {\bibfnamefont {P.}~\bibnamefont {Sengupta}}, \bibinfo {author} {\bibfnamefont {C.}~\bibnamefont {Kloc}},\ and\ \bibinfo {author} {\bibfnamefont {Q.}~\bibnamefont {Xiong}},\ }\bibfield  {title} {\bibinfo {title} {{Raman spectroscopy of atomically thin two-dimensional magnetic iron phosphorus trisulfide (FePS$_3$) crystals}},\ }\href {https://doi.org/10.1088/2053-1583/3/3/031009} {\bibfield  {journal} {\bibinfo  {journal}
  {2D Materials}\ }\textbf {\bibinfo {volume} {3}},\ \bibinfo {pages} {031009} (\bibinfo {year} {2016})}\BibitemShut {NoStop}%
\bibitem [{\citenamefont {Kim}\ \emph {et~al.}(2019{\natexlab{c}})\citenamefont {Kim}, \citenamefont {Lee},\ and\ \citenamefont {Cheong}}]{Kim_rev}%
  \BibitemOpen
  \bibfield  {author} {\bibinfo {author} {\bibfnamefont {K.}~\bibnamefont {Kim}}, \bibinfo {author} {\bibfnamefont {J.-U.}\ \bibnamefont {Lee}},\ and\ \bibinfo {author} {\bibfnamefont {H.}~\bibnamefont {Cheong}},\ }\bibfield  {title} {\bibinfo {title} {{Raman spectroscopy of two-dimensional magnetic van der Waals materials}},\ }\href {https://doi.org/10.1088/1361-6528/ab37a4} {\bibfield  {journal} {\bibinfo  {journal} {Nanotechnology}\ }\textbf {\bibinfo {volume} {30}},\ \bibinfo {pages} {452001} (\bibinfo {year} {2019}{\natexlab{c}})}\BibitemShut {NoStop}%
\bibitem [{\citenamefont {Zhang}\ \emph {et~al.}(2020)\citenamefont {Zhang}, \citenamefont {Wu}, \citenamefont {Lyu}, \citenamefont {Wu}, \citenamefont {Zhao}, \citenamefont {Chen}, \citenamefont {Jia}, \citenamefont {Zhang}, \citenamefont {Wang}, \citenamefont {Wang}, \citenamefont {Chen}, \citenamefont {Mei}, \citenamefont {Taniguchi}, \citenamefont {Watanabe}, \citenamefont {Yan}, \citenamefont {Liu}, \citenamefont {Huang}, \citenamefont {Zhao},\ and\ \citenamefont {Huang}}]{Zhang2020}%
  \BibitemOpen
  \bibfield  {author} {\bibinfo {author} {\bibfnamefont {Y.}~\bibnamefont {Zhang}}, \bibinfo {author} {\bibfnamefont {X.}~\bibnamefont {Wu}}, \bibinfo {author} {\bibfnamefont {B.}~\bibnamefont {Lyu}}, \bibinfo {author} {\bibfnamefont {M.}~\bibnamefont {Wu}}, \bibinfo {author} {\bibfnamefont {S.}~\bibnamefont {Zhao}}, \bibinfo {author} {\bibfnamefont {J.}~\bibnamefont {Chen}}, \bibinfo {author} {\bibfnamefont {M.}~\bibnamefont {Jia}}, \bibinfo {author} {\bibfnamefont {C.}~\bibnamefont {Zhang}}, \bibinfo {author} {\bibfnamefont {L.}~\bibnamefont {Wang}}, \bibinfo {author} {\bibfnamefont {X.}~\bibnamefont {Wang}}, \bibinfo {author} {\bibfnamefont {Y.}~\bibnamefont {Chen}}, \bibinfo {author} {\bibfnamefont {J.}~\bibnamefont {Mei}}, \bibinfo {author} {\bibfnamefont {T.}~\bibnamefont {Taniguchi}}, \bibinfo {author} {\bibfnamefont {K.}~\bibnamefont {Watanabe}}, \bibinfo {author} {\bibfnamefont {H.}~\bibnamefont {Yan}}, \bibinfo {author} {\bibfnamefont {Q.}~\bibnamefont {Liu}}, \bibinfo {author} {\bibfnamefont
  {L.}~\bibnamefont {Huang}}, \bibinfo {author} {\bibfnamefont {Y.}~\bibnamefont {Zhao}},\ and\ \bibinfo {author} {\bibfnamefont {M.}~\bibnamefont {Huang}},\ }\bibfield  {title} {\bibinfo {title} {{Magnetic Order-Induced Polarization Anomaly of Raman Scattering in 2D Magnet CrI$_3$}},\ }\href {https://doi.org/10.1021/acs.nanolett.9b04634} {\bibfield  {journal} {\bibinfo  {journal} {Nano Letters}\ }\textbf {\bibinfo {volume} {20}},\ \bibinfo {pages} {729} (\bibinfo {year} {2020})}\BibitemShut {NoStop}%
\bibitem [{\citenamefont {Li}\ \emph {et~al.}(2020)\citenamefont {Li}, \citenamefont {Ye}, \citenamefont {Luo}, \citenamefont {Ye}, \citenamefont {Kim}, \citenamefont {Yang}, \citenamefont {Tian}, \citenamefont {Li}, \citenamefont {Lei}, \citenamefont {Tsen}, \citenamefont {Sun}, \citenamefont {He},\ and\ \citenamefont {Zhao}}]{PhysRevX.10.011075}%
  \BibitemOpen
  \bibfield  {author} {\bibinfo {author} {\bibfnamefont {S.}~\bibnamefont {Li}}, \bibinfo {author} {\bibfnamefont {Z.}~\bibnamefont {Ye}}, \bibinfo {author} {\bibfnamefont {X.}~\bibnamefont {Luo}}, \bibinfo {author} {\bibfnamefont {G.}~\bibnamefont {Ye}}, \bibinfo {author} {\bibfnamefont {H.~H.}\ \bibnamefont {Kim}}, \bibinfo {author} {\bibfnamefont {B.}~\bibnamefont {Yang}}, \bibinfo {author} {\bibfnamefont {S.}~\bibnamefont {Tian}}, \bibinfo {author} {\bibfnamefont {C.}~\bibnamefont {Li}}, \bibinfo {author} {\bibfnamefont {H.}~\bibnamefont {Lei}}, \bibinfo {author} {\bibfnamefont {A.~W.}\ \bibnamefont {Tsen}}, \bibinfo {author} {\bibfnamefont {K.}~\bibnamefont {Sun}}, \bibinfo {author} {\bibfnamefont {R.}~\bibnamefont {He}},\ and\ \bibinfo {author} {\bibfnamefont {L.}~\bibnamefont {Zhao}},\ }\bibfield  {title} {\bibinfo {title} {{Magnetic-Field-Induced Quantum Phase Transitions in a van der Waals Magnet}},\ }\href {https://doi.org/10.1103/PhysRevX.10.011075} {\bibfield  {journal} {\bibinfo  {journal} {Phys.
  Rev. X}\ }\textbf {\bibinfo {volume} {10}},\ \bibinfo {pages} {011075} (\bibinfo {year} {2020})}\BibitemShut {NoStop}%
\bibitem [{\citenamefont {Huang}\ \emph {et~al.}(2020)\citenamefont {Huang}, \citenamefont {Cenker}, \citenamefont {Zhang}, \citenamefont {Ray}, \citenamefont {Song}, \citenamefont {Taniguchi}, \citenamefont {Watanabe}, \citenamefont {McGuire}, \citenamefont {Xiao},\ and\ \citenamefont {Xu}}]{Huang2020}%
  \BibitemOpen
  \bibfield  {author} {\bibinfo {author} {\bibfnamefont {B.}~\bibnamefont {Huang}}, \bibinfo {author} {\bibfnamefont {J.}~\bibnamefont {Cenker}}, \bibinfo {author} {\bibfnamefont {X.}~\bibnamefont {Zhang}}, \bibinfo {author} {\bibfnamefont {E.~L.}\ \bibnamefont {Ray}}, \bibinfo {author} {\bibfnamefont {T.}~\bibnamefont {Song}}, \bibinfo {author} {\bibfnamefont {T.}~\bibnamefont {Taniguchi}}, \bibinfo {author} {\bibfnamefont {K.}~\bibnamefont {Watanabe}}, \bibinfo {author} {\bibfnamefont {M.~A.}\ \bibnamefont {McGuire}}, \bibinfo {author} {\bibfnamefont {D.}~\bibnamefont {Xiao}},\ and\ \bibinfo {author} {\bibfnamefont {X.}~\bibnamefont {Xu}},\ }\bibfield  {title} {\bibinfo {title} {{Tuning inelastic light scattering via symmetry control in the two-dimensional magnet CrI$_3$}},\ }\href {https://doi.org/10.1038/s41565-019-0598-4} {\bibfield  {journal} {\bibinfo  {journal} {Nature Nanotechnology}\ }\textbf {\bibinfo {volume} {15}},\ \bibinfo {pages} {212} (\bibinfo {year} {2020})}\BibitemShut {NoStop}%
\bibitem [{\citenamefont {McCreary}\ \emph {et~al.}(2020)\citenamefont {McCreary}, \citenamefont {Mai}, \citenamefont {Utermohlen}, \citenamefont {Simpson}, \citenamefont {Garrity}, \citenamefont {Feng}, \citenamefont {Shcherbakov}, \citenamefont {Zhu}, \citenamefont {Hu}, \citenamefont {Weber}, \citenamefont {Watanabe}, \citenamefont {Taniguchi}, \citenamefont {Goldberger}, \citenamefont {Mao}, \citenamefont {Lau}, \citenamefont {Lu}, \citenamefont {Trivedi}, \citenamefont {Vald{\'e}s~Aguilar},\ and\ \citenamefont {Hight~Walker}}]{McCreary2020}%
  \BibitemOpen
  \bibfield  {author} {\bibinfo {author} {\bibfnamefont {A.}~\bibnamefont {McCreary}}, \bibinfo {author} {\bibfnamefont {T.~T.}\ \bibnamefont {Mai}}, \bibinfo {author} {\bibfnamefont {F.~G.}\ \bibnamefont {Utermohlen}}, \bibinfo {author} {\bibfnamefont {J.~R.}\ \bibnamefont {Simpson}}, \bibinfo {author} {\bibfnamefont {K.~F.}\ \bibnamefont {Garrity}}, \bibinfo {author} {\bibfnamefont {X.}~\bibnamefont {Feng}}, \bibinfo {author} {\bibfnamefont {D.}~\bibnamefont {Shcherbakov}}, \bibinfo {author} {\bibfnamefont {Y.}~\bibnamefont {Zhu}}, \bibinfo {author} {\bibfnamefont {J.}~\bibnamefont {Hu}}, \bibinfo {author} {\bibfnamefont {D.}~\bibnamefont {Weber}}, \bibinfo {author} {\bibfnamefont {K.}~\bibnamefont {Watanabe}}, \bibinfo {author} {\bibfnamefont {T.}~\bibnamefont {Taniguchi}}, \bibinfo {author} {\bibfnamefont {J.~E.}\ \bibnamefont {Goldberger}}, \bibinfo {author} {\bibfnamefont {Z.}~\bibnamefont {Mao}}, \bibinfo {author} {\bibfnamefont {C.~N.}\ \bibnamefont {Lau}}, \bibinfo {author} {\bibfnamefont
  {Y.}~\bibnamefont {Lu}}, \bibinfo {author} {\bibfnamefont {N.}~\bibnamefont {Trivedi}}, \bibinfo {author} {\bibfnamefont {R.}~\bibnamefont {Vald{\'e}s~Aguilar}},\ and\ \bibinfo {author} {\bibfnamefont {A.~R.}\ \bibnamefont {Hight~Walker}},\ }\bibfield  {title} {\bibinfo {title} {{Distinct magneto-Raman signatures of spin-flip phase transitions in CrI$_3$}},\ }\href {https://doi.org/10.1038/s41467-020-17320-3} {\bibfield  {journal} {\bibinfo  {journal} {Nature Communications}\ }\textbf {\bibinfo {volume} {11}},\ \bibinfo {pages} {3879} (\bibinfo {year} {2020})}\BibitemShut {NoStop}%
\bibitem [{\citenamefont {Lyu}\ \emph {et~al.}(2020)\citenamefont {Lyu}, \citenamefont {Gao}, \citenamefont {Zhang}, \citenamefont {Wang}, \citenamefont {Wu}, \citenamefont {Chen}, \citenamefont {Zhang}, \citenamefont {Li}, \citenamefont {Huang}, \citenamefont {Zhang}, \citenamefont {Chen}, \citenamefont {Mei}, \citenamefont {Yan}, \citenamefont {Zhao}, \citenamefont {Huang},\ and\ \citenamefont {Huang}}]{Lyu2020}%
  \BibitemOpen
  \bibfield  {author} {\bibinfo {author} {\bibfnamefont {B.}~\bibnamefont {Lyu}}, \bibinfo {author} {\bibfnamefont {Y.}~\bibnamefont {Gao}}, \bibinfo {author} {\bibfnamefont {Y.}~\bibnamefont {Zhang}}, \bibinfo {author} {\bibfnamefont {L.}~\bibnamefont {Wang}}, \bibinfo {author} {\bibfnamefont {X.}~\bibnamefont {Wu}}, \bibinfo {author} {\bibfnamefont {Y.}~\bibnamefont {Chen}}, \bibinfo {author} {\bibfnamefont {J.}~\bibnamefont {Zhang}}, \bibinfo {author} {\bibfnamefont {G.}~\bibnamefont {Li}}, \bibinfo {author} {\bibfnamefont {Q.}~\bibnamefont {Huang}}, \bibinfo {author} {\bibfnamefont {N.}~\bibnamefont {Zhang}}, \bibinfo {author} {\bibfnamefont {Y.}~\bibnamefont {Chen}}, \bibinfo {author} {\bibfnamefont {J.}~\bibnamefont {Mei}}, \bibinfo {author} {\bibfnamefont {H.}~\bibnamefont {Yan}}, \bibinfo {author} {\bibfnamefont {Y.}~\bibnamefont {Zhao}}, \bibinfo {author} {\bibfnamefont {L.}~\bibnamefont {Huang}},\ and\ \bibinfo {author} {\bibfnamefont {M.}~\bibnamefont {Huang}},\ }\bibfield  {title} {\bibinfo
  {title} {{Probing the Ferromagnetism and Spin Wave Gap in VI$_3$ by Helicity-Resolved Raman Spectroscopy}},\ }\href {https://doi.org/10.1021/acs.nanolett.0c02029} {\bibfield  {journal} {\bibinfo  {journal} {Nano Letters}\ }\textbf {\bibinfo {volume} {20}},\ \bibinfo {pages} {6024} (\bibinfo {year} {2020})}\BibitemShut {NoStop}%
\bibitem [{\citenamefont {Cenker}\ \emph {et~al.}(2021)\citenamefont {Cenker}, \citenamefont {Huang}, \citenamefont {Suri}, \citenamefont {Thijssen}, \citenamefont {Miller}, \citenamefont {Song}, \citenamefont {Taniguchi}, \citenamefont {Watanabe}, \citenamefont {McGuire}, \citenamefont {Xiao},\ and\ \citenamefont {Xu}}]{Cenker2021}%
  \BibitemOpen
  \bibfield  {author} {\bibinfo {author} {\bibfnamefont {J.}~\bibnamefont {Cenker}}, \bibinfo {author} {\bibfnamefont {B.}~\bibnamefont {Huang}}, \bibinfo {author} {\bibfnamefont {N.}~\bibnamefont {Suri}}, \bibinfo {author} {\bibfnamefont {P.}~\bibnamefont {Thijssen}}, \bibinfo {author} {\bibfnamefont {A.}~\bibnamefont {Miller}}, \bibinfo {author} {\bibfnamefont {T.}~\bibnamefont {Song}}, \bibinfo {author} {\bibfnamefont {T.}~\bibnamefont {Taniguchi}}, \bibinfo {author} {\bibfnamefont {K.}~\bibnamefont {Watanabe}}, \bibinfo {author} {\bibfnamefont {M.~A.}\ \bibnamefont {McGuire}}, \bibinfo {author} {\bibfnamefont {D.}~\bibnamefont {Xiao}},\ and\ \bibinfo {author} {\bibfnamefont {X.}~\bibnamefont {Xu}},\ }\bibfield  {title} {\bibinfo {title} {{Direct observation of two-dimensional magnons in atomically thin CrI$_3$}},\ }\href {https://doi.org/10.1038/s41567-020-0999-1} {\bibfield  {journal} {\bibinfo  {journal} {Nature Physics}\ }\textbf {\bibinfo {volume} {17}},\ \bibinfo {pages} {20} (\bibinfo {year}
  {2021})}\BibitemShut {NoStop}%
\bibitem [{\citenamefont {Zhang}\ and\ \citenamefont {Niu}(2015)}]{PhysRevLett.115.115502}%
  \BibitemOpen
  \bibfield  {author} {\bibinfo {author} {\bibfnamefont {L.}~\bibnamefont {Zhang}}\ and\ \bibinfo {author} {\bibfnamefont {Q.}~\bibnamefont {Niu}},\ }\bibfield  {title} {\bibinfo {title} {{Chiral Phonons at High-Symmetry Points in Monolayer Hexagonal Lattices}},\ }\href {https://doi.org/10.1103/PhysRevLett.115.115502} {\bibfield  {journal} {\bibinfo  {journal} {Phys. Rev. Lett.}\ }\textbf {\bibinfo {volume} {115}},\ \bibinfo {pages} {115502} (\bibinfo {year} {2015})}\BibitemShut {NoStop}%
\bibitem [{\citenamefont {Zhu}\ \emph {et~al.}(2018)\citenamefont {Zhu}, \citenamefont {Yi}, \citenamefont {Li}, \citenamefont {Xiao}, \citenamefont {Zhang}, \citenamefont {Yang}, \citenamefont {Kaindl}, \citenamefont {Li}, \citenamefont {Wang},\ and\ \citenamefont {Zhang}}]{Zhu_Science2018}%
  \BibitemOpen
  \bibfield  {author} {\bibinfo {author} {\bibfnamefont {H.}~\bibnamefont {Zhu}}, \bibinfo {author} {\bibfnamefont {J.}~\bibnamefont {Yi}}, \bibinfo {author} {\bibfnamefont {M.-Y.}\ \bibnamefont {Li}}, \bibinfo {author} {\bibfnamefont {J.}~\bibnamefont {Xiao}}, \bibinfo {author} {\bibfnamefont {L.}~\bibnamefont {Zhang}}, \bibinfo {author} {\bibfnamefont {C.-W.}\ \bibnamefont {Yang}}, \bibinfo {author} {\bibfnamefont {R.~A.}\ \bibnamefont {Kaindl}}, \bibinfo {author} {\bibfnamefont {L.-J.}\ \bibnamefont {Li}}, \bibinfo {author} {\bibfnamefont {Y.}~\bibnamefont {Wang}},\ and\ \bibinfo {author} {\bibfnamefont {X.}~\bibnamefont {Zhang}},\ }\bibfield  {title} {\bibinfo {title} {{Observation of chiral phonons}},\ }\href {https://doi.org/10.1126/science.aar2711} {\bibfield  {journal} {\bibinfo  {journal} {Science}\ }\textbf {\bibinfo {volume} {359}},\ \bibinfo {pages} {579} (\bibinfo {year} {2018})}\BibitemShut {NoStop}%
\bibitem [{\citenamefont {Zhang}\ and\ \citenamefont {Murakami}(2022)}]{zhang2022chiral}%
  \BibitemOpen
  \bibfield  {author} {\bibinfo {author} {\bibfnamefont {T.}~\bibnamefont {Zhang}}\ and\ \bibinfo {author} {\bibfnamefont {S.}~\bibnamefont {Murakami}},\ }\bibfield  {title} {\bibinfo {title} {Chiral phonons and pseudoangular momentum in nonsymmorphic systems},\ }\href@noop {} {\bibfield  {journal} {\bibinfo  {journal} {Physical Review Research}\ }\textbf {\bibinfo {volume} {4}},\ \bibinfo {pages} {L012024} (\bibinfo {year} {2022})}\BibitemShut {NoStop}%
\bibitem [{\citenamefont {Ueda}\ \emph {et~al.}(2023)\citenamefont {Ueda}, \citenamefont {Garc{\'i}a-Fern{\'a}ndez}, \citenamefont {Agrestini}, \citenamefont {Romao}, \citenamefont {van~den Brink}, \citenamefont {Spaldin}, \citenamefont {Zhou},\ and\ \citenamefont {Staub}}]{Ueda2023}%
  \BibitemOpen
  \bibfield  {author} {\bibinfo {author} {\bibfnamefont {H.}~\bibnamefont {Ueda}}, \bibinfo {author} {\bibfnamefont {M.}~\bibnamefont {Garc{\'i}a-Fern{\'a}ndez}}, \bibinfo {author} {\bibfnamefont {S.}~\bibnamefont {Agrestini}}, \bibinfo {author} {\bibfnamefont {C.~P.}\ \bibnamefont {Romao}}, \bibinfo {author} {\bibfnamefont {J.}~\bibnamefont {van~den Brink}}, \bibinfo {author} {\bibfnamefont {N.~A.}\ \bibnamefont {Spaldin}}, \bibinfo {author} {\bibfnamefont {K.-J.}\ \bibnamefont {Zhou}},\ and\ \bibinfo {author} {\bibfnamefont {U.}~\bibnamefont {Staub}},\ }\bibfield  {title} {\bibinfo {title} {{Chiral phonons in quartz probed by X-rays}},\ }\href {https://doi.org/10.1038/s41586-023-06016-5} {\bibfield  {journal} {\bibinfo  {journal} {Nature}\ }\textbf {\bibinfo {volume} {618}},\ \bibinfo {pages} {946} (\bibinfo {year} {2023})}\BibitemShut {NoStop}%
\bibitem [{\citenamefont {Ishito}\ \emph {et~al.}(2023)\citenamefont {Ishito}, \citenamefont {Mao}, \citenamefont {Kousaka}, \citenamefont {Togawa}, \citenamefont {Iwasaki}, \citenamefont {Zhang}, \citenamefont {Murakami}, \citenamefont {Kishine},\ and\ \citenamefont {Satoh}}]{Ishito2023}%
  \BibitemOpen
  \bibfield  {author} {\bibinfo {author} {\bibfnamefont {K.}~\bibnamefont {Ishito}}, \bibinfo {author} {\bibfnamefont {H.}~\bibnamefont {Mao}}, \bibinfo {author} {\bibfnamefont {Y.}~\bibnamefont {Kousaka}}, \bibinfo {author} {\bibfnamefont {Y.}~\bibnamefont {Togawa}}, \bibinfo {author} {\bibfnamefont {S.}~\bibnamefont {Iwasaki}}, \bibinfo {author} {\bibfnamefont {T.}~\bibnamefont {Zhang}}, \bibinfo {author} {\bibfnamefont {S.}~\bibnamefont {Murakami}}, \bibinfo {author} {\bibfnamefont {J.-i.}\ \bibnamefont {Kishine}},\ and\ \bibinfo {author} {\bibfnamefont {T.}~\bibnamefont {Satoh}},\ }\bibfield  {title} {\bibinfo {title} {{Truly chiral phonons in $\alpha$-HgS}},\ }\href {https://doi.org/10.1038/s41567-022-01790-x} {\bibfield  {journal} {\bibinfo  {journal} {Nature Physics}\ }\textbf {\bibinfo {volume} {19}},\ \bibinfo {pages} {35} (\bibinfo {year} {2023})}\BibitemShut {NoStop}%
\bibitem [{\citenamefont {Zhang}\ \emph {et~al.}(2023)\citenamefont {Zhang}, \citenamefont {Huang}, \citenamefont {Pan}, \citenamefont {Du}, \citenamefont {Zhang},\ and\ \citenamefont {Murakami}}]{zhang2023weyl}%
  \BibitemOpen
  \bibfield  {author} {\bibinfo {author} {\bibfnamefont {T.}~\bibnamefont {Zhang}}, \bibinfo {author} {\bibfnamefont {Z.}~\bibnamefont {Huang}}, \bibinfo {author} {\bibfnamefont {Z.}~\bibnamefont {Pan}}, \bibinfo {author} {\bibfnamefont {L.}~\bibnamefont {Du}}, \bibinfo {author} {\bibfnamefont {G.}~\bibnamefont {Zhang}},\ and\ \bibinfo {author} {\bibfnamefont {S.}~\bibnamefont {Murakami}},\ }\bibfield  {title} {\bibinfo {title} {Weyl phonons in chiral crystals},\ }\href@noop {} {\bibfield  {journal} {\bibinfo  {journal} {Nano Letters}\ }\textbf {\bibinfo {volume} {23}},\ \bibinfo {pages} {7561} (\bibinfo {year} {2023})}\BibitemShut {NoStop}%
\bibitem [{\citenamefont {Zhang}\ \emph {et~al.}(2026)\citenamefont {Zhang}, \citenamefont {Huang}, \citenamefont {Du}, \citenamefont {Ying}, \citenamefont {Du},\ and\ \citenamefont {Zhang}}]{zhang2025chirality}%
  \BibitemOpen
  \bibfield  {author} {\bibinfo {author} {\bibfnamefont {S.}~\bibnamefont {Zhang}}, \bibinfo {author} {\bibfnamefont {Z.}~\bibnamefont {Huang}}, \bibinfo {author} {\bibfnamefont {M.}~\bibnamefont {Du}}, \bibinfo {author} {\bibfnamefont {T.}~\bibnamefont {Ying}}, \bibinfo {author} {\bibfnamefont {L.}~\bibnamefont {Du}},\ and\ \bibinfo {author} {\bibfnamefont {T.}~\bibnamefont {Zhang}},\ }\bibfield  {title} {\bibinfo {title} {Comprehensive study of phonon chirality under symmetry constraints},\ }\href@noop {} {\bibfield  {journal} {\bibinfo  {journal} {Physical Review B}\ }\textbf {\bibinfo {volume} {113}},\ \bibinfo {pages} {024302} (\bibinfo {year} {2026})}\BibitemShut {NoStop}%
\bibitem [{\citenamefont {Zhang}\ \emph {et~al.}(2025{\natexlab{a}})\citenamefont {Zhang}, \citenamefont {Liu}, \citenamefont {Miao},\ and\ \citenamefont {Murakami}}]{zhang2025advances}%
  \BibitemOpen
  \bibfield  {author} {\bibinfo {author} {\bibfnamefont {T.}~\bibnamefont {Zhang}}, \bibinfo {author} {\bibfnamefont {Y.}~\bibnamefont {Liu}}, \bibinfo {author} {\bibfnamefont {H.}~\bibnamefont {Miao}},\ and\ \bibinfo {author} {\bibfnamefont {S.}~\bibnamefont {Murakami}},\ }\bibfield  {title} {\bibinfo {title} {Advances in phonons: From band topology to phonon chirality},\ }\href@noop {} {\bibfield  {journal} {\bibinfo  {journal} {arXiv preprint arXiv:2505.06179}\ } (\bibinfo {year} {2025}{\natexlab{a}})}\BibitemShut {NoStop}%
\bibitem [{\citenamefont {Zhang}\ and\ \citenamefont {Niu}(2014)}]{ZhangNiu_PRL2014}%
  \BibitemOpen
  \bibfield  {author} {\bibinfo {author} {\bibfnamefont {L.}~\bibnamefont {Zhang}}\ and\ \bibinfo {author} {\bibfnamefont {Q.}~\bibnamefont {Niu}},\ }\bibfield  {title} {\bibinfo {title} {{Angular Momentum of Phonons and the Einstein--de Haas Effect}},\ }\href {https://doi.org/10.1103/PhysRevLett.112.085503} {\bibfield  {journal} {\bibinfo  {journal} {Phys. Rev. Lett.}\ }\textbf {\bibinfo {volume} {112}},\ \bibinfo {pages} {085503} (\bibinfo {year} {2014})}\BibitemShut {NoStop}%
\bibitem [{\citenamefont {Jin}\ \emph {et~al.}(2020)\citenamefont {Jin}, \citenamefont {Ye}, \citenamefont {Luo}, \citenamefont {Yang}, \citenamefont {Ye}, \citenamefont {Yin}, \citenamefont {Kim}, \citenamefont {Rojas}, \citenamefont {Tian}, \citenamefont {Fu}, \citenamefont {Yan}, \citenamefont {Lei}, \citenamefont {Sun}, \citenamefont {Tsen}, \citenamefont {He},\ and\ \citenamefont {Zhao}}]{Jin_PNAS2020}%
  \BibitemOpen
  \bibfield  {author} {\bibinfo {author} {\bibfnamefont {W.}~\bibnamefont {Jin}}, \bibinfo {author} {\bibfnamefont {Z.}~\bibnamefont {Ye}}, \bibinfo {author} {\bibfnamefont {X.}~\bibnamefont {Luo}}, \bibinfo {author} {\bibfnamefont {B.}~\bibnamefont {Yang}}, \bibinfo {author} {\bibfnamefont {G.}~\bibnamefont {Ye}}, \bibinfo {author} {\bibfnamefont {F.}~\bibnamefont {Yin}}, \bibinfo {author} {\bibfnamefont {H.~H.}\ \bibnamefont {Kim}}, \bibinfo {author} {\bibfnamefont {L.}~\bibnamefont {Rojas}}, \bibinfo {author} {\bibfnamefont {S.}~\bibnamefont {Tian}}, \bibinfo {author} {\bibfnamefont {Y.}~\bibnamefont {Fu}}, \bibinfo {author} {\bibfnamefont {S.}~\bibnamefont {Yan}}, \bibinfo {author} {\bibfnamefont {H.}~\bibnamefont {Lei}}, \bibinfo {author} {\bibfnamefont {K.}~\bibnamefont {Sun}}, \bibinfo {author} {\bibfnamefont {A.~W.}\ \bibnamefont {Tsen}}, \bibinfo {author} {\bibfnamefont {R.}~\bibnamefont {He}},\ and\ \bibinfo {author} {\bibfnamefont {L.}~\bibnamefont {Zhao}},\ }\bibfield  {title} {\bibinfo {title}
  {{Tunable layered-magnetism-assisted magneto-Raman effect in a two-dimensional magnet CrI$_{3}$}},\ }\href {https://doi.org/10.1073/pnas.2012980117} {\bibfield  {journal} {\bibinfo  {journal} {Proceedings of the National Academy of Sciences}\ }\textbf {\bibinfo {volume} {117}},\ \bibinfo {pages} {24664} (\bibinfo {year} {2020})}\BibitemShut {NoStop}%
\bibitem [{\citenamefont {Hernandez}\ \emph {et~al.}(2023)\citenamefont {Hernandez}, \citenamefont {Baydin}, \citenamefont {Chaudhary}, \citenamefont {Tay}, \citenamefont {Katayama}, \citenamefont {Takeda}, \citenamefont {Nojiri}, \citenamefont {Okazaki}, \citenamefont {Rappl}, \citenamefont {Abramof}, \citenamefont {Rodriguez-Vega}, \citenamefont {Fiete},\ and\ \citenamefont {Kono}}]{Hernandez_SciAdv2023}%
  \BibitemOpen
  \bibfield  {author} {\bibinfo {author} {\bibfnamefont {F.~G.~G.}\ \bibnamefont {Hernandez}}, \bibinfo {author} {\bibfnamefont {A.}~\bibnamefont {Baydin}}, \bibinfo {author} {\bibfnamefont {S.}~\bibnamefont {Chaudhary}}, \bibinfo {author} {\bibfnamefont {F.}~\bibnamefont {Tay}}, \bibinfo {author} {\bibfnamefont {I.}~\bibnamefont {Katayama}}, \bibinfo {author} {\bibfnamefont {J.}~\bibnamefont {Takeda}}, \bibinfo {author} {\bibfnamefont {H.}~\bibnamefont {Nojiri}}, \bibinfo {author} {\bibfnamefont {A.~K.}\ \bibnamefont {Okazaki}}, \bibinfo {author} {\bibfnamefont {P.~H.~O.}\ \bibnamefont {Rappl}}, \bibinfo {author} {\bibfnamefont {E.}~\bibnamefont {Abramof}}, \bibinfo {author} {\bibfnamefont {M.}~\bibnamefont {Rodriguez-Vega}}, \bibinfo {author} {\bibfnamefont {G.~A.}\ \bibnamefont {Fiete}},\ and\ \bibinfo {author} {\bibfnamefont {J.}~\bibnamefont {Kono}},\ }\bibfield  {title} {\bibinfo {title} {{Observation of interplay between phonon chirality and electronic band topology}},\ }\href
  {https://doi.org/10.1126/sciadv.adj4074} {\bibfield  {journal} {\bibinfo  {journal} {Science Advances}\ }\textbf {\bibinfo {volume} {9}},\ \bibinfo {pages} {eadj4074} (\bibinfo {year} {2023})}\BibitemShut {NoStop}%
\bibitem [{\citenamefont {Yang}\ \emph {et~al.}(2025)\citenamefont {Yang}, \citenamefont {Zhu}, \citenamefont {Steigleder}, \citenamefont {Liu}, \citenamefont {Liu}, \citenamefont {Qiu}, \citenamefont {Zhang},\ and\ \citenamefont {Dressel}}]{PhysRevLett.134.196905}%
  \BibitemOpen
  \bibfield  {author} {\bibinfo {author} {\bibfnamefont {R.}~\bibnamefont {Yang}}, \bibinfo {author} {\bibfnamefont {Y.-Y.}\ \bibnamefont {Zhu}}, \bibinfo {author} {\bibfnamefont {M.}~\bibnamefont {Steigleder}}, \bibinfo {author} {\bibfnamefont {Y.-C.}\ \bibnamefont {Liu}}, \bibinfo {author} {\bibfnamefont {C.-C.}\ \bibnamefont {Liu}}, \bibinfo {author} {\bibfnamefont {X.-G.}\ \bibnamefont {Qiu}}, \bibinfo {author} {\bibfnamefont {T.}~\bibnamefont {Zhang}},\ and\ \bibinfo {author} {\bibfnamefont {M.}~\bibnamefont {Dressel}},\ }\bibfield  {title} {\bibinfo {title} {{Inherent Circular Dichroism of Phonons in Magnetic Weyl Semimetal ${\mathrm{Co}}_{3}{\text{Sn}}_{2}{\mathrm{S}}_{2}$}},\ }\href {https://doi.org/10.1103/PhysRevLett.134.196905} {\bibfield  {journal} {\bibinfo  {journal} {Phys. Rev. Lett.}\ }\textbf {\bibinfo {volume} {134}},\ \bibinfo {pages} {196905} (\bibinfo {year} {2025})}\BibitemShut {NoStop}%
\bibitem [{\citenamefont {Che}\ \emph {et~al.}(2025)\citenamefont {Che}, \citenamefont {Liang}, \citenamefont {Cui}, \citenamefont {Li}, \citenamefont {Lu}, \citenamefont {Sang}, \citenamefont {Li}, \citenamefont {Dong}, \citenamefont {Zhao}, \citenamefont {Zhang}, \citenamefont {Sun}, \citenamefont {Jiang}, \citenamefont {Liu}, \citenamefont {Jin}, \citenamefont {Zhang},\ and\ \citenamefont {Yang}}]{PhysRevLett.134.196906}%
  \BibitemOpen
  \bibfield  {author} {\bibinfo {author} {\bibfnamefont {M.}~\bibnamefont {Che}}, \bibinfo {author} {\bibfnamefont {J.}~\bibnamefont {Liang}}, \bibinfo {author} {\bibfnamefont {Y.}~\bibnamefont {Cui}}, \bibinfo {author} {\bibfnamefont {H.}~\bibnamefont {Li}}, \bibinfo {author} {\bibfnamefont {B.}~\bibnamefont {Lu}}, \bibinfo {author} {\bibfnamefont {W.}~\bibnamefont {Sang}}, \bibinfo {author} {\bibfnamefont {X.}~\bibnamefont {Li}}, \bibinfo {author} {\bibfnamefont {X.}~\bibnamefont {Dong}}, \bibinfo {author} {\bibfnamefont {L.}~\bibnamefont {Zhao}}, \bibinfo {author} {\bibfnamefont {S.}~\bibnamefont {Zhang}}, \bibinfo {author} {\bibfnamefont {T.}~\bibnamefont {Sun}}, \bibinfo {author} {\bibfnamefont {W.}~\bibnamefont {Jiang}}, \bibinfo {author} {\bibfnamefont {E.}~\bibnamefont {Liu}}, \bibinfo {author} {\bibfnamefont {F.}~\bibnamefont {Jin}}, \bibinfo {author} {\bibfnamefont {T.}~\bibnamefont {Zhang}},\ and\ \bibinfo {author} {\bibfnamefont {L.}~\bibnamefont {Yang}},\ }\bibfield  {title} {\bibinfo {title}
  {{Magnetic Order Induced Chiral Phonons in a Ferromagnetic Weyl Semimetal}},\ }\href {https://doi.org/10.1103/PhysRevLett.134.196906} {\bibfield  {journal} {\bibinfo  {journal} {Phys. Rev. Lett.}\ }\textbf {\bibinfo {volume} {134}},\ \bibinfo {pages} {196906} (\bibinfo {year} {2025})}\BibitemShut {NoStop}%
\bibitem [{\citenamefont {Lukachuk}\ \emph {et~al.}(2007)\citenamefont {Lukachuk}, \citenamefont {Zheng}, \citenamefont {Mattausch}, \citenamefont {Kremer}, \citenamefont {Simon},\ and\ \citenamefont {Banks}}]{ggi0}%
  \BibitemOpen
  \bibfield  {author} {\bibinfo {author} {\bibfnamefont {M.}~\bibnamefont {Lukachuk}}, \bibinfo {author} {\bibfnamefont {C.}~\bibnamefont {Zheng}}, \bibinfo {author} {\bibfnamefont {H.}~\bibnamefont {Mattausch}}, \bibinfo {author} {\bibfnamefont {R.~K.}\ \bibnamefont {Kremer}}, \bibinfo {author} {\bibfnamefont {A.}~\bibnamefont {Simon}},\ and\ \bibinfo {author} {\bibfnamefont {M.~G.}\ \bibnamefont {Banks}},\ }\bibfield  {title} {\bibinfo {title} {{RE$_{2+x}$I$_2$M$_{2+y}$ (RE = Ce,Gd, Y; M = Al, Ga): Reduced Rare Earth Halides with a Hexagonal Metal Atom Network}},\ }\href {https://doi.org/doi:10.1515/znb-2007-0502} {\bibfield  {journal} {\bibinfo  {journal} {Zeitschrift für Naturforschung B}\ }\textbf {\bibinfo {volume} {62}},\ \bibinfo {pages} {633} (\bibinfo {year} {2007})}\BibitemShut {NoStop}%
\bibitem [{\citenamefont {Kaneko}\ \emph {et~al.}(2026)\citenamefont {Kaneko}, \citenamefont {Mizuno}, \citenamefont {Kamiyama}, \citenamefont {Miyamoto},\ and\ \citenamefont {Ochi}}]{ggikaneko}%
  \BibitemOpen
  \bibfield  {author} {\bibinfo {author} {\bibfnamefont {T.}~\bibnamefont {Kaneko}}, \bibinfo {author} {\bibfnamefont {R.}~\bibnamefont {Mizuno}}, \bibinfo {author} {\bibfnamefont {S.}~\bibnamefont {Kamiyama}}, \bibinfo {author} {\bibfnamefont {H.}~\bibnamefont {Miyamoto}},\ and\ \bibinfo {author} {\bibfnamefont {M.}~\bibnamefont {Ochi}},\ }\bibfield  {title} {\bibinfo {title} {Electronic band structure, phonon dispersion, and magnetic triple-$q$ state in gdgai},\ }\href {https://doi.org/10.1103/pt6l-xvlx} {\bibfield  {journal} {\bibinfo  {journal} {Phys. Rev. B}\ }\textbf {\bibinfo {volume} {113}},\ \bibinfo {pages} {045156} (\bibinfo {year} {2026})}\BibitemShut {NoStop}%
\bibitem [{\citenamefont {Okuma}\ \emph {et~al.}(2024)\citenamefont {Okuma}, \citenamefont {Yamagami}, \citenamefont {Fujisawa}, \citenamefont {Hsu}, \citenamefont {Obata}, \citenamefont {Tomoda}, \citenamefont {Dronova}, \citenamefont {Kuroda}, \citenamefont {Ishikawa}, \citenamefont {Kawaguchi}, \citenamefont {Aido}, \citenamefont {Kindo}, \citenamefont {Chan}, \citenamefont {Lin}, \citenamefont {Ihara}, \citenamefont {Kondo},\ and\ \citenamefont {Okada}}]{okumaggi}%
  \BibitemOpen
  \bibfield  {author} {\bibinfo {author} {\bibfnamefont {R.}~\bibnamefont {Okuma}}, \bibinfo {author} {\bibfnamefont {K.}~\bibnamefont {Yamagami}}, \bibinfo {author} {\bibfnamefont {Y.}~\bibnamefont {Fujisawa}}, \bibinfo {author} {\bibfnamefont {C.~H.}\ \bibnamefont {Hsu}}, \bibinfo {author} {\bibfnamefont {Y.}~\bibnamefont {Obata}}, \bibinfo {author} {\bibfnamefont {N.}~\bibnamefont {Tomoda}}, \bibinfo {author} {\bibfnamefont {M.}~\bibnamefont {Dronova}}, \bibinfo {author} {\bibfnamefont {K.}~\bibnamefont {Kuroda}}, \bibinfo {author} {\bibfnamefont {H.}~\bibnamefont {Ishikawa}}, \bibinfo {author} {\bibfnamefont {K.}~\bibnamefont {Kawaguchi}}, \bibinfo {author} {\bibfnamefont {K.}~\bibnamefont {Aido}}, \bibinfo {author} {\bibfnamefont {K.}~\bibnamefont {Kindo}}, \bibinfo {author} {\bibfnamefont {Y.~H.}\ \bibnamefont {Chan}}, \bibinfo {author} {\bibfnamefont {H.}~\bibnamefont {Lin}}, \bibinfo {author} {\bibfnamefont {Y.}~\bibnamefont {Ihara}}, \bibinfo {author} {\bibfnamefont {T.}~\bibnamefont {Kondo}},\ and\
  \bibinfo {author} {\bibfnamefont {Y.}~\bibnamefont {Okada}},\ }\href {https://arxiv.org/abs/2405.16781} {\bibinfo {title} {Emergent topological magnetism in hund's excitonic insulator}} (\bibinfo {year} {2024}),\ \Eprint {https://arxiv.org/abs/2405.16781} {arXiv:2405.16781 [cond-mat.str-el]} \BibitemShut {NoStop}%
\bibitem [{\citenamefont {Zhang}\ \emph {et~al.}(2022)\citenamefont {Zhang}, \citenamefont {Zhao}, \citenamefont {Wang},\ and\ \citenamefont {Smet}}]{Zhang2022}%
  \BibitemOpen
  \bibfield  {author} {\bibinfo {author} {\bibfnamefont {Y.}~\bibnamefont {Zhang}}, \bibinfo {author} {\bibfnamefont {D.}~\bibnamefont {Zhao}}, \bibinfo {author} {\bibfnamefont {Q.}~\bibnamefont {Wang}},\ and\ \bibinfo {author} {\bibfnamefont {J.~H.}\ \bibnamefont {Smet}},\ }\bibfield  {title} {\bibinfo {title} {{In situ Raman spectroscopy across superconducting transition of liquid-gated MoS$_2$}},\ }\href@noop {} {\bibfield  {journal} {\bibinfo  {journal} {Applied Physics Letters}\ }\textbf {\bibinfo {volume} {120}},\ \bibinfo {pages} {053106} (\bibinfo {year} {2022})}\BibitemShut {NoStop}%
\bibitem [{\citenamefont {Sun}\ \emph {et~al.}(2021)\citenamefont {Sun}, \citenamefont {Pang},\ and\ \citenamefont {Zhang}}]{sun2021review}%
  \BibitemOpen
  \bibfield  {author} {\bibinfo {author} {\bibfnamefont {Y.-J.}\ \bibnamefont {Sun}}, \bibinfo {author} {\bibfnamefont {S.-M.}\ \bibnamefont {Pang}},\ and\ \bibinfo {author} {\bibfnamefont {J.}~\bibnamefont {Zhang}},\ }\bibfield  {title} {\bibinfo {title} {Review of raman spectroscopy of two-dimensional magnetic van der waals materials},\ }\href@noop {} {\bibfield  {journal} {\bibinfo  {journal} {Chinese Physics B}\ }\textbf {\bibinfo {volume} {30}},\ \bibinfo {pages} {117104} (\bibinfo {year} {2021})}\BibitemShut {NoStop}%
\bibitem [{\citenamefont {Wang}\ \emph {et~al.}(2024)\citenamefont {Wang}, \citenamefont {Sun}, \citenamefont {Li},\ and\ \citenamefont {Zhang}}]{wang2024chiral}%
  \BibitemOpen
  \bibfield  {author} {\bibinfo {author} {\bibfnamefont {T.}~\bibnamefont {Wang}}, \bibinfo {author} {\bibfnamefont {H.}~\bibnamefont {Sun}}, \bibinfo {author} {\bibfnamefont {X.}~\bibnamefont {Li}},\ and\ \bibinfo {author} {\bibfnamefont {L.}~\bibnamefont {Zhang}},\ }\bibfield  {title} {\bibinfo {title} {Chiral phonons: Prediction, verification, and application},\ }\href@noop {} {\bibfield  {journal} {\bibinfo  {journal} {Nano Letters}\ }\textbf {\bibinfo {volume} {24}},\ \bibinfo {pages} {4311} (\bibinfo {year} {2024})}\BibitemShut {NoStop}%
\bibitem [{\citenamefont {Zhang}\ \emph {et~al.}(2025{\natexlab{b}})\citenamefont {Zhang}, \citenamefont {Wang},\ and\ \citenamefont {Zhang}}]{zhang2025general}%
  \BibitemOpen
  \bibfield  {author} {\bibinfo {author} {\bibfnamefont {S.}~\bibnamefont {Zhang}}, \bibinfo {author} {\bibfnamefont {M.}~\bibnamefont {Wang}},\ and\ \bibinfo {author} {\bibfnamefont {T.}~\bibnamefont {Zhang}},\ }\bibfield  {title} {\bibinfo {title} {General ab initio framework for chiral phonons induced by electronic order},\ }\href@noop {} {\bibfield  {journal} {\bibinfo  {journal} {arXiv preprint arXiv:2509.09253}\ } (\bibinfo {year} {2025}{\natexlab{b}})}\BibitemShut {NoStop}%
\bibitem [{\citenamefont {Juraschek}\ \emph {et~al.}(2025)\citenamefont {Juraschek}, \citenamefont {Geilhufe}, \citenamefont {Zhu}, \citenamefont {Basini}, \citenamefont {Baum}, \citenamefont {Baydin}, \citenamefont {Chaudhary}, \citenamefont {Fechner}, \citenamefont {Flebus}, \citenamefont {Grissonnanche} \emph {et~al.}}]{juraschek2025chiral}%
  \BibitemOpen
  \bibfield  {author} {\bibinfo {author} {\bibfnamefont {D.~M.}\ \bibnamefont {Juraschek}}, \bibinfo {author} {\bibfnamefont {R.~M.}\ \bibnamefont {Geilhufe}}, \bibinfo {author} {\bibfnamefont {H.}~\bibnamefont {Zhu}}, \bibinfo {author} {\bibfnamefont {M.}~\bibnamefont {Basini}}, \bibinfo {author} {\bibfnamefont {P.}~\bibnamefont {Baum}}, \bibinfo {author} {\bibfnamefont {A.}~\bibnamefont {Baydin}}, \bibinfo {author} {\bibfnamefont {S.}~\bibnamefont {Chaudhary}}, \bibinfo {author} {\bibfnamefont {M.}~\bibnamefont {Fechner}}, \bibinfo {author} {\bibfnamefont {B.}~\bibnamefont {Flebus}}, \bibinfo {author} {\bibfnamefont {G.}~\bibnamefont {Grissonnanche}}, \emph {et~al.},\ }\bibfield  {title} {\bibinfo {title} {Chiral phonons},\ }\href@noop {} {\bibfield  {journal} {\bibinfo  {journal} {Nature Physics}\ }\textbf {\bibinfo {volume} {21}},\ \bibinfo {pages} {1532} (\bibinfo {year} {2025})}\BibitemShut {NoStop}%
\bibitem{SM}
See Supplemental Material at [URL will be inserted by publisher] for details of sample preparation, x-ray diffraction measurements, device fabrication, polarized Raman measurements, first-principles calculations, room-temperature polarized Raman spectra, Raman-intensity maps, Raman-active modes at the M point, lattice constants, magnetic-field-dependent Raman spectra, and reproducibility measurements. The Supplemental Material also contains Refs.~\cite{kresse1996efficient,perdew1996generalized,togo2023implementation}.
\bibitem [{\citenamefont {Kresse}\ and\ \citenamefont {Furthm{\"u}ller}(1996)}]{kresse1996efficient}%
  \BibitemOpen
  \bibfield  {author} {\bibinfo {author} {\bibfnamefont {G.}~\bibnamefont {Kresse}}\ and\ \bibinfo {author} {\bibfnamefont {J.}~\bibnamefont {Furthm{\"u}ller}},\ }\bibfield  {title} {\bibinfo {title} {Efficient iterative schemes for ab initio total-energy calculations using a plane-wave basis set},\ }\href@noop {} {\bibfield  {journal} {\bibinfo  {journal} {Physical review B}\ }\textbf {\bibinfo {volume} {54}},\ \bibinfo {pages} {11169} (\bibinfo {year} {1996})}\BibitemShut {NoStop}%
\bibitem [{\citenamefont {Perdew}\ \emph {et~al.}(1996)\citenamefont {Perdew}, \citenamefont {Burke},\ and\ \citenamefont {Ernzerhof}}]{perdew1996generalized}%
  \BibitemOpen
  \bibfield  {author} {\bibinfo {author} {\bibfnamefont {J.~P.}\ \bibnamefont {Perdew}}, \bibinfo {author} {\bibfnamefont {K.}~\bibnamefont {Burke}},\ and\ \bibinfo {author} {\bibfnamefont {M.}~\bibnamefont {Ernzerhof}},\ }\bibfield  {title} {\bibinfo {title} {Generalized gradient approximation made simple},\ }\href@noop {} {\bibfield  {journal} {\bibinfo  {journal} {Physical review letters}\ }\textbf {\bibinfo {volume} {77}},\ \bibinfo {pages} {3865} (\bibinfo {year} {1996})}\BibitemShut {NoStop}%
\bibitem [{\citenamefont {Togo}\ \emph {et~al.}(2023)\citenamefont {Togo}, \citenamefont {Chaput}, \citenamefont {Tadano},\ and\ \citenamefont {Tanaka}}]{togo2023implementation}%
  \BibitemOpen
  \bibfield  {author} {\bibinfo {author} {\bibfnamefont {A.}~\bibnamefont {Togo}}, \bibinfo {author} {\bibfnamefont {L.}~\bibnamefont {Chaput}}, \bibinfo {author} {\bibfnamefont {T.}~\bibnamefont {Tadano}},\ and\ \bibinfo {author} {\bibfnamefont {I.}~\bibnamefont {Tanaka}},\ }\bibfield  {title} {\bibinfo {title} {Implementation strategies in phonopy and phono3py},\ }\href@noop {} {\bibfield  {journal} {\bibinfo  {journal} {Journal of Physics: Condensed Matter}\ }\textbf {\bibinfo {volume} {35}},\ \bibinfo {pages} {353001} (\bibinfo {year} {2023})}\BibitemShut {NoStop}%
\bibitem [{\citenamefont {Loudon}(1964)}]{Loudon01101964}%
  \BibitemOpen
  \bibfield  {author} {\bibinfo {author} {\bibfnamefont {R.}~\bibnamefont {Loudon}},\ }\bibfield  {title} {\bibinfo {title} {{The Raman effect in crystals}},\ }\href {https://doi.org/10.1080/00018736400101051} {\bibfield  {journal} {\bibinfo  {journal} {Advances in Physics}\ }\textbf {\bibinfo {volume} {13}},\ \bibinfo {pages} {423} (\bibinfo {year} {1964})}\BibitemShut {NoStop}%
\bibitem [{\citenamefont {Men\'endez}\ and\ \citenamefont {Cardona}(1984)}]{PhysRevB.29.2051}%
  \BibitemOpen
  \bibfield  {author} {\bibinfo {author} {\bibfnamefont {J.}~\bibnamefont {Men\'endez}}\ and\ \bibinfo {author} {\bibfnamefont {M.}~\bibnamefont {Cardona}},\ }\bibfield  {title} {\bibinfo {title} {{Temperature dependence of the first-order Raman scattering by phonons in Si, Ge, and $\ensuremath{\alpha}\ensuremath{-}\mathrm{S}\mathrm{n}$: Anharmonic effects}},\ }\href {https://doi.org/10.1103/PhysRevB.29.2051} {\bibfield  {journal} {\bibinfo  {journal} {Phys. Rev. B}\ }\textbf {\bibinfo {volume} {29}},\ \bibinfo {pages} {2051} (\bibinfo {year} {1984})}\BibitemShut {NoStop}%
\bibitem [{\citenamefont {Irmer}\ \emph {et~al.}(1996)\citenamefont {Irmer}, \citenamefont {Wenzel},\ and\ \citenamefont {Monecke}}]{Irmer1996}%
  \BibitemOpen
  \bibfield  {author} {\bibinfo {author} {\bibfnamefont {G.}~\bibnamefont {Irmer}}, \bibinfo {author} {\bibfnamefont {M.}~\bibnamefont {Wenzel}},\ and\ \bibinfo {author} {\bibfnamefont {J.}~\bibnamefont {Monecke}},\ }\bibfield  {title} {\bibinfo {title} {{The temperature dependence of the LO(T) and TO(T) phonons in GaAs and InP}},\ }\href {https://doi.org/https://doi.org/10.1002/pssb.2221950110} {\bibfield  {journal} {\bibinfo  {journal} {physica status solidi (b)}\ }\textbf {\bibinfo {volume} {195}},\ \bibinfo {pages} {85} (\bibinfo {year} {1996})}\BibitemShut {NoStop}%
\bibitem [{\citenamefont {Link}\ \emph {et~al.}(1999)\citenamefont {Link}, \citenamefont {Bitzer}, \citenamefont {Limmer}, \citenamefont {Sauer}, \citenamefont {Kirchner}, \citenamefont {Schwegler}, \citenamefont {Kamp}, \citenamefont {Ebling},\ and\ \citenamefont {Benz}}]{Link1999}%
  \BibitemOpen
  \bibfield  {author} {\bibinfo {author} {\bibfnamefont {A.}~\bibnamefont {Link}}, \bibinfo {author} {\bibfnamefont {K.}~\bibnamefont {Bitzer}}, \bibinfo {author} {\bibfnamefont {W.}~\bibnamefont {Limmer}}, \bibinfo {author} {\bibfnamefont {R.}~\bibnamefont {Sauer}}, \bibinfo {author} {\bibfnamefont {C.}~\bibnamefont {Kirchner}}, \bibinfo {author} {\bibfnamefont {V.}~\bibnamefont {Schwegler}}, \bibinfo {author} {\bibfnamefont {M.}~\bibnamefont {Kamp}}, \bibinfo {author} {\bibfnamefont {D.~G.}\ \bibnamefont {Ebling}},\ and\ \bibinfo {author} {\bibfnamefont {K.~W.}\ \bibnamefont {Benz}},\ }\bibfield  {title} {\bibinfo {title} {{Temperature dependence of the E2 and A1(LO) phonons in GaN and AlN}},\ }\href {https://doi.org/10.1063/1.371681} {\bibfield  {journal} {\bibinfo  {journal} {Journal of Applied Physics}\ }\textbf {\bibinfo {volume} {86}},\ \bibinfo {pages} {6256} (\bibinfo {year} {1999})}\BibitemShut {NoStop}%
\bibitem [{\citenamefont {Munn}(1975)}]{PhysRevB.12.3491}%
  \BibitemOpen
  \bibfield  {author} {\bibinfo {author} {\bibfnamefont {R.~W.}\ \bibnamefont {Munn}},\ }\bibfield  {title} {\bibinfo {title} {Gr\"uneisen parameters for molecular crystals},\ }\href {https://doi.org/10.1103/PhysRevB.12.3491} {\bibfield  {journal} {\bibinfo  {journal} {Phys. Rev. B}\ }\textbf {\bibinfo {volume} {12}},\ \bibinfo {pages} {3491} (\bibinfo {year} {1975})}\BibitemShut {NoStop}%
\bibitem [{\citenamefont {Ritz}\ \emph {et~al.}(2019)\citenamefont {Ritz}, \citenamefont {Li},\ and\ \citenamefont {Benedek}}]{Ethan2019}%
  \BibitemOpen
  \bibfield  {author} {\bibinfo {author} {\bibfnamefont {E.~T.}\ \bibnamefont {Ritz}}, \bibinfo {author} {\bibfnamefont {S.~J.}\ \bibnamefont {Li}},\ and\ \bibinfo {author} {\bibfnamefont {N.~A.}\ \bibnamefont {Benedek}},\ }\bibfield  {title} {\bibinfo {title} {Thermal expansion in insulating solids from first principles},\ }\href {https://doi.org/10.1063/1.5125779} {\bibfield  {journal} {\bibinfo  {journal} {Journal of Applied Physics}\ }\textbf {\bibinfo {volume} {126}},\ \bibinfo {pages} {171102} (\bibinfo {year} {2019})}\BibitemShut {NoStop}%
\bibitem [{\citenamefont {Yin}\ \emph {et~al.}(2021)\citenamefont {Yin}, \citenamefont {Ulman}, \citenamefont {Liu}, \citenamefont {Granados~del Águila}, \citenamefont {Huang}, \citenamefont {Zhang}, \citenamefont {Serra}, \citenamefont {Sedmidubsky}, \citenamefont {Sofer}, \citenamefont {Quek},\ and\ \citenamefont {Xiong}}]{adma.202101618}%
  \BibitemOpen
  \bibfield  {author} {\bibinfo {author} {\bibfnamefont {T.}~\bibnamefont {Yin}}, \bibinfo {author} {\bibfnamefont {K.~A.}\ \bibnamefont {Ulman}}, \bibinfo {author} {\bibfnamefont {S.}~\bibnamefont {Liu}}, \bibinfo {author} {\bibfnamefont {A.}~\bibnamefont {Granados~del Águila}}, \bibinfo {author} {\bibfnamefont {Y.}~\bibnamefont {Huang}}, \bibinfo {author} {\bibfnamefont {L.}~\bibnamefont {Zhang}}, \bibinfo {author} {\bibfnamefont {M.}~\bibnamefont {Serra}}, \bibinfo {author} {\bibfnamefont {D.}~\bibnamefont {Sedmidubsky}}, \bibinfo {author} {\bibfnamefont {Z.}~\bibnamefont {Sofer}}, \bibinfo {author} {\bibfnamefont {S.~Y.}\ \bibnamefont {Quek}},\ and\ \bibinfo {author} {\bibfnamefont {Q.}~\bibnamefont {Xiong}},\ }\bibfield  {title} {\bibinfo {title} {{Chiral Phonons and Giant Magneto-Optical Effect in CrBr3 2D Magnet}},\ }\href {https://doi.org/https://doi.org/10.1002/adma.202101618} {\bibfield  {journal} {\bibinfo  {journal} {Advanced Materials}\ }\textbf {\bibinfo {volume} {33}},\ \bibinfo {pages}
  {2101618} (\bibinfo {year} {2021})}\BibitemShut {NoStop}%
\bibitem [{\citenamefont {Liu}\ \emph {et~al.}(2023)\citenamefont {Liu}, \citenamefont {Long},\ and\ \citenamefont {Wang}}]{PhysRevB.108.184414}%
  \BibitemOpen
  \bibfield  {author} {\bibinfo {author} {\bibfnamefont {S.}~\bibnamefont {Liu}}, \bibinfo {author} {\bibfnamefont {M.-Q.}\ \bibnamefont {Long}},\ and\ \bibinfo {author} {\bibfnamefont {Y.-P.}\ \bibnamefont {Wang}},\ }\bibfield  {title} {\bibinfo {title} {{Theoretical investigations on the magneto-Raman effect of ${\text{CrI}}_{3}$}},\ }\href {https://doi.org/10.1103/PhysRevB.108.184414} {\bibfield  {journal} {\bibinfo  {journal} {Phys. Rev. B}\ }\textbf {\bibinfo {volume} {108}},\ \bibinfo {pages} {184414} (\bibinfo {year} {2023})}\BibitemShut {NoStop}%
\bibitem [{\citenamefont {Rao}\ \emph {et~al.}(2025)\citenamefont {Rao}, \citenamefont {Jiang}, \citenamefont {Pachter}, \citenamefont {Mai}, \citenamefont {Mohaugen}, \citenamefont {Muñoz}, \citenamefont {Siebenaller}, \citenamefont {Rowe}, \citenamefont {Selhorst}, \citenamefont {Giordano}, \citenamefont {Walker},\ and\ \citenamefont {Susner}}]{rao2025}%
  \BibitemOpen
  \bibfield  {author} {\bibinfo {author} {\bibfnamefont {R.}~\bibnamefont {Rao}}, \bibinfo {author} {\bibfnamefont {J.}~\bibnamefont {Jiang}}, \bibinfo {author} {\bibfnamefont {R.}~\bibnamefont {Pachter}}, \bibinfo {author} {\bibfnamefont {T.~T.}\ \bibnamefont {Mai}}, \bibinfo {author} {\bibfnamefont {V.}~\bibnamefont {Mohaugen}}, \bibinfo {author} {\bibfnamefont {M.~F.}\ \bibnamefont {Muñoz}}, \bibinfo {author} {\bibfnamefont {R.}~\bibnamefont {Siebenaller}}, \bibinfo {author} {\bibfnamefont {E.}~\bibnamefont {Rowe}}, \bibinfo {author} {\bibfnamefont {R.}~\bibnamefont {Selhorst}}, \bibinfo {author} {\bibfnamefont {A.~N.}\ \bibnamefont {Giordano}}, \bibinfo {author} {\bibfnamefont {A.~R.~H.}\ \bibnamefont {Walker}},\ and\ \bibinfo {author} {\bibfnamefont {M.~A.}\ \bibnamefont {Susner}},\ }\href {https://arxiv.org/abs/2501.17565} {\bibinfo {title} {{Anomalous Raman scattering in layered AgCrP$_2$Se$_6$: Helical modes and excitation energy-dependent intensities}}} (\bibinfo {year} {2025}),\ \Eprint
  {https://arxiv.org/abs/2501.17565} {arXiv:2501.17565 [cond-mat.mes-hall]} \BibitemShut {NoStop}%
\bibitem [{\citenamefont {Anastassakis}\ \emph {et~al.}(1972)\citenamefont {Anastassakis}, \citenamefont {Burstein}, \citenamefont {Maradudin},\ and\ \citenamefont {Minnick}}]{ANASTASSAKIS1972519}%
  \BibitemOpen
  \bibfield  {author} {\bibinfo {author} {\bibfnamefont {E.}~\bibnamefont {Anastassakis}}, \bibinfo {author} {\bibfnamefont {E.}~\bibnamefont {Burstein}}, \bibinfo {author} {\bibfnamefont {A.}~\bibnamefont {Maradudin}},\ and\ \bibinfo {author} {\bibfnamefont {R.}~\bibnamefont {Minnick}},\ }\bibfield  {title} {\bibinfo {title} {{Morphic effects — III. Effects of an external magnetic field on the long wavelength optical phonons}},\ }\href {https://doi.org/https://doi.org/10.1016/0022-3697(72)90034-0} {\bibfield  {journal} {\bibinfo  {journal} {Journal of Physics and Chemistry of Solids}\ }\textbf {\bibinfo {volume} {33}},\ \bibinfo {pages} {519} (\bibinfo {year} {1972})}\BibitemShut {NoStop}%
\bibitem [{\citenamefont {Momma}\ and\ \citenamefont {Izumi}(2011)}]{vesta}%
  \BibitemOpen
  \bibfield  {author} {\bibinfo {author} {\bibfnamefont {K.}~\bibnamefont {Momma}}\ and\ \bibinfo {author} {\bibfnamefont {F.}~\bibnamefont {Izumi}},\ }\bibfield  {title} {\bibinfo {title} {{VESTA 3 for three-dimensional visualization of crystal, volumetric and morphology data}},\ }\href@noop {} {\bibfield  {journal} {\bibinfo  {journal} {Journal of Applied Crystallography}\ }\textbf {\bibinfo {volume} {44}},\ \bibinfo {pages} {1272} (\bibinfo {year} {2011})}\BibitemShut {NoStop}%
\end{thebibliography}
%

\end{document}